\documentclass[aps,pre,onecolumn,showpacs,a4paper]{revtex4}
\usepackage{graphicx} 
\usepackage{amsmath}
\usepackage{amssymb}
\usepackage{amscd}

\begin{document}

\title{Anomalous diffusion in nonlinear oscillators with multiplicative noise}
\author{Kirone Mallick}
 \affiliation{Service de Physique Th\'eorique, Centre d'\'Etudes de Saclay,
 91191 Gif-sur-Yvette Cedex, France}
 \email{mallick@spht.saclay.cea.fr}
\author{Philippe Marcq}
 \affiliation{Institut de Recherche sur les Ph\'enom\`enes Hors \'Equilibre,
 Universit\'e de Provence,
 49 rue Joliot-Curie, BP 146, 13384 Marseille Cedex 13, France}
 \email{marcq@irphe.univ-mrs.fr}
\date{April 18, 2002}
\begin{abstract}
The time-asymptotic behavior of undamped, nonlinear oscillators with 
a random frequency is investigated analytically and numerically. 
We find that averaged quantities of physical interest, such as 
the oscillator's mechanical energy, root-mean-square position and 
velocity, grow algebraically with time. The scaling exponents and 
associated generalized diffusion constants are calculated when the 
oscillator's potential energy grows as  a power of its position: 
${\mathcal U}(x) \sim x^{2 n}$ for $|x| \rightarrow \infty$. 
Correlated noise yields anomalous diffusion exponents equal
to half the value found for white noise.
\end{abstract}
\pacs{05.10.Gg,05.40.-a,05.45.-a}
\maketitle 

 \section{Introduction}

  Randomness in the external conditions  entails  the  parameters of a
   dynamical system to fluctuate.  The extent  of these fluctuations
   is independent  of  any  thermodynamic 
  characteristic of the system  in contrast  to 
   intrinsic fluctuations  the  amplitude of which is  proportional  to 
 the equilibrium  temperature,  in accordance with 
  the fluctuation-dissipation  theorem \cite{vankampen,risken}. 
  Usually,    external randomness appears  as a 
  multiplicative  noise  in the dynamical equations.
 The interplay  of noise and nonlinearity in a system
 far from equilibrium   results  in some unusual  phenomena \cite{lefever}.
  In fact,  the  presence of noise   dramatically   alters  
 the  properties of a nonlinear  dynamical system  both 
 qualitatively  and   quantitatively 
 (for a recent review see \cite{landaMc}).
 For example, it was shown recently
 that in a spatially-extended system, a 
 multiplicative noise,  white in space and time,
 generates   an ordered symmetry-breaking state through
 a non-equilibrium phase transition  whereas no such transition
 exists in the absence  of noise \cite{vandenb1,vandenb2}.
 Noise  can  also  induce  spatial patterns 
 \cite{vandenb3,toral}  or  improve the performance
  of a nonlinear device through   stochastic resonance 
 \cite{barbay}. Furthermore, even  if some 
  important   qualitative  features of a   deterministic system
    survive to   external noise,
 their  quantitative characteristics  may  change:
 a stable fixed point may become unstable \cite{bourret}, 
  a bifurcation  may  be delayed 
  (noise-induced stabilization) \cite{lucke,roder} and 
 scale-invariant properties  which manifest themselves as power-laws 
  may be altered with  the appearance of  non-classical  scaling exponents
 \cite{genovese}.

  The discovery of  Brownian motors which are able
 to rectify random fluctuations into a  directed 
 motion (noise-induced transport)
 has triggered a renewed interest  in the study of simple  one-dimensional
 mechanical  models of   particles in a
  potential with   random parameters \cite{reimann}.  
   It is well known that a   linear oscillator 
  submitted to   parametric  noise
  can be unstable  even if damping is taken into account
 \cite{bourret,linden}. This noise-induced  energetic instability
  has been observed in  diverse 
  experimental contexts such as electronic oscillators
 \cite{strato,kabashima}, nematic liquid crystals \cite{kawakubo}
 and  surface waves (Faraday  instability)  \cite{fauve}.
 In engineering fields this instability
 plays a crucial role in the study of  the dynamic response of flexible
 structures to random environmental loading such as the wave-induced
 motion of off-shore structures or the vibration of tall buildings
 in a turbulent wind \cite{roberts}. The presence of nonlinear
  friction tends   to limit the oscillation amplitude:  the
  pendulum with randomly vibrating suspension axis and undergoing
  nonlinear friction,  known as the van der Pol oscillator, has
 been studied   in the small noise limit using perturbative
 expansions \cite{strato,landa}. 

  In the present work, we  consider   the 
 motion of an undamped nonlinear oscillator trapped in a general
 confining potential and 
 submitted to   parametric random fluctuations. 
 Because there is no  dissipation, the energy of  the  system
 increases with  time and 
 we shall show  that  the position, 
 the momentum and  the energy grow as  
 power-laws of time with  nontrivial
  exponents  that depend on the behavior of the
 confining potential at infinity \cite{bouchaud}.  A  key  feature
 of our method  is to use the integrability
 properties of the associated deterministic
 nonlinear oscillator in order  to derive exact
 stochastic equations in action-angle variables.
 We  then  use  
 the averaging  technique  of classical mechanics \cite{landau},
 together with a reduction procedure \cite{vandenbroeck1,drolet},
  to  calculate exactly the anomalous scaling 
  exponents, irrespective of  the amplitude of the noise. 
  Some of our results were derived before in the particular case of
  a cubic nonlinearity using an energy-enveloppe equation \cite{lind2}. 
 Our method enables us to derive  the numerical prefactors 
 appearing in the scaling laws
  (generalized diffusion constants), and our  analytical
 predictions  compare  very satisfactorily  with the 
 numerical results. In the case of  noise   correlated in time,
 the anomalous diffusion exponents are modified:  they can be 
   obtained by dimensional analysis arguments and the values
 thus found also agree with numerical results.
 Throughout  this work,  crossover phenomena  between different 
 scaling regimes are emphasized.

 This article is organized as follows. In section \ref{sec:linear}, we recall 
 that   the energy of a  linear oscillator with multiplicative noise 
 grows exponentially with time and that  this growth may be characterized
 by a Lyapunov exponent. In section \ref{sec:duff}, 
we analyze  the  classical   Duffing 
 oscillator   in presence of parametric  noise. Our technique
 allows us  to study  precisely the long-time behavior 
 of the system.    In section \ref{sec:gen},
 we consider  a particle in an arbitrary confining
 potential that grows as a polynomial at large distances.
 In section \ref{sec:color}, we discuss  the case of  colored 
 noise where the presence  of  a new timescale (the correlation time)  
 leads to  a nontrivial crossover from the white noise regime to 
 another scaling regime.
  Our conclusions are presented in section \ref{sec:conc}. 
  In  Appendix \ref{sec:add}, 
  the nonlinear oscillator in presence of both additive and
 multiplicative noise is  briefly studied: we show that
 at long times the effect of additive  noise  is irrelevant.
 Appendix \ref{sec:num} is devoted to numerical 
 methods and  in Appendix \ref{sec:elliptic} 
 some  useful mathematical relations are recalled.

\section{The linear oscillator with parametric noise}
\label{sec:linear}

  In this section  we recall known results 
 for an undamped linear oscillator   submitted to  parametric
 noise, a  generic and widely studied  model,  in order to understand
 the role of external multiplicative noise \cite{lefever,bourret,strato}.
 The dynamical equation  for such a system is 
\begin{equation}
   \frac{\textrm{d}^2 }{\textrm{d} t^2}x(t) +(\omega^2 + \xi(t))\, x(t) = 0 ,
 \label{multlin}
\end{equation}
 where $x(t)$ represents the position of the oscillator at time $t$ and 
 $\omega$  its frequency. The random   noise  $\xi(t)$ 
 is a Gaussian white noise  of zero mean-value and 
  of amplitude  ${\mathcal D}$:
 \begin{eqnarray}
       \langle \xi(t)  \rangle &=&   0   \, ,\nonumber \\
   \langle \xi(t) \xi(t') \rangle  &=&  {\mathcal D} \, \delta( t - t') .
   \label{defgamma}
 \end{eqnarray} 
 
The physical interpretation of Eq.~(\ref{multlin})  is that 
 the frequency of the
 oscillator is not constant in time but fluctuates around its
 mean value $\omega$ because of randomness in the external
 conditions (external noise). 
 When   these fluctuations  are   deterministic and 
 periodic  in time, Eq.~(\ref{multlin}) is a Mathieu equation
 which has been extensively studied \cite{landau}. 
 Here,  we are interested in   the case  where  these fluctuations
  are   random  with no  deterministic part.
  The origin, $x = 0$ and $ \textrm{d}x/\textrm{d}t = 0$, is an
  unstable stationary solution of  Eq.~(\ref{multlin}).
 As  shown in Ref.~\cite{bourret},
 this instability can be studied  from 
  the dynamical   evolution of the  Probability Distribution Function
 $P(x,v,t)$ of $x$ and $v$ (with $v = \dot x =  \textrm{d}x/\textrm{d}t)$. 
  This P.D.F. obeys  the Fokker-Planck equation \cite{vankampen,risken}
  associated with  Eq.~(\ref{multlin}):
 \begin{equation}
\frac{ \partial P}{\partial t} = - v\frac{\partial P}{\partial x}
 + \omega^2 x \frac{ \partial P}{\partial v}
 +\frac{{\mathcal D}}{2}\frac{ \partial^2}{\partial v^2}\left( x^2 P \right),
\label{FPlin}
\end{equation}
 where Eq.~(\ref{multlin}) is understood according to Stratonovich rules.

 This  Fokker-Planck   equation   leads to a closed
 system of  ordinary differential equations that  couple 
 the $n +1 $    moments of order $n$,
 {\it i.e.},  moments  of the type   $\langle x^{n-k} v^{k} \rangle$,
 where $n$  and $k$ are  positive integers and 
 $ 0 \le k \le n$:
   \begin{equation}
 \frac{\textrm{d}}{\textrm{d} t} \langle x^{n-k} v^{k} \rangle
   =  (n-k) \, \langle x^{n-k-1} v^{k+1} \rangle -  \omega^2 k
     \, \langle x^{n-k+1} v^{k-1} \rangle +  \frac{\mathcal D}{2} 
     \, k(k-1)\,  \langle x^{n-k +2} v^{k-2} \rangle .
 \label{linsyst}
    \end{equation}
 The divergence of the  moments with time 
 results from the existence of at least one 
 positive eigenvalue of the linear system (\ref{linsyst}).
  In particular,  the mean value of the mechanical
 energy $E$   of the system ({\it i.e.} the sum of 
 its kinetic and   potential energies) grows exponentially  with time: 
\begin{equation}
 \langle E  \rangle =  \frac{1}{2} \, \langle v^2 \rangle + 
   \frac{1}{2} \omega^2 \, \langle  x^2 \rangle  \propto  e^{\mu \, t} ,
\label{moyenerg}
\end{equation}
 where  the growth rate  $\mu$ is the positive real root of
 the equation:
\begin{equation}
 \mu^3 + 4 \omega^2 \mu = 2{\mathcal D}  .
\label{taux}
 \end{equation}

\begin{figure}[ht]
\hfill
\includegraphics*[width=0.4\textwidth]{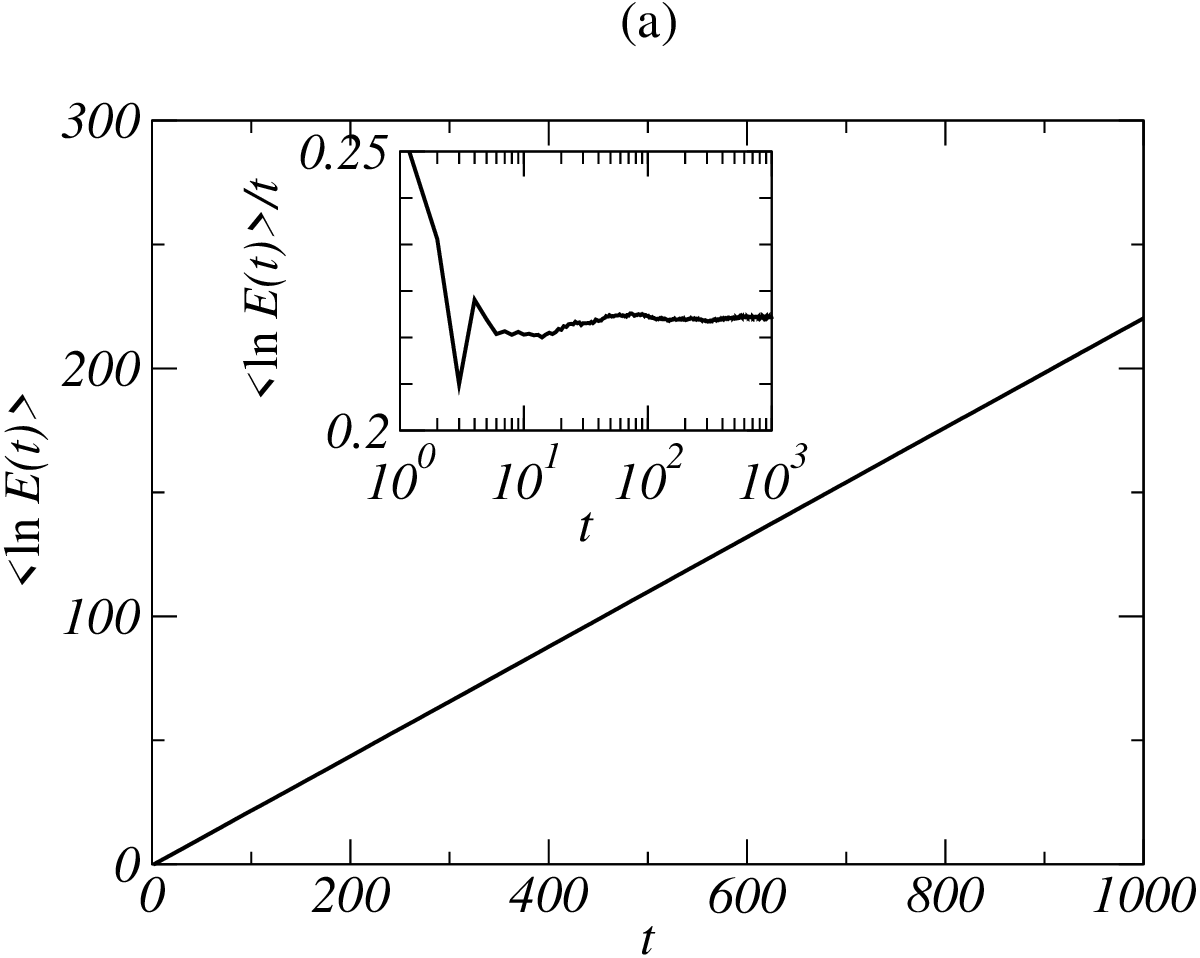}
\hfill
\includegraphics*[width=0.4\textwidth]{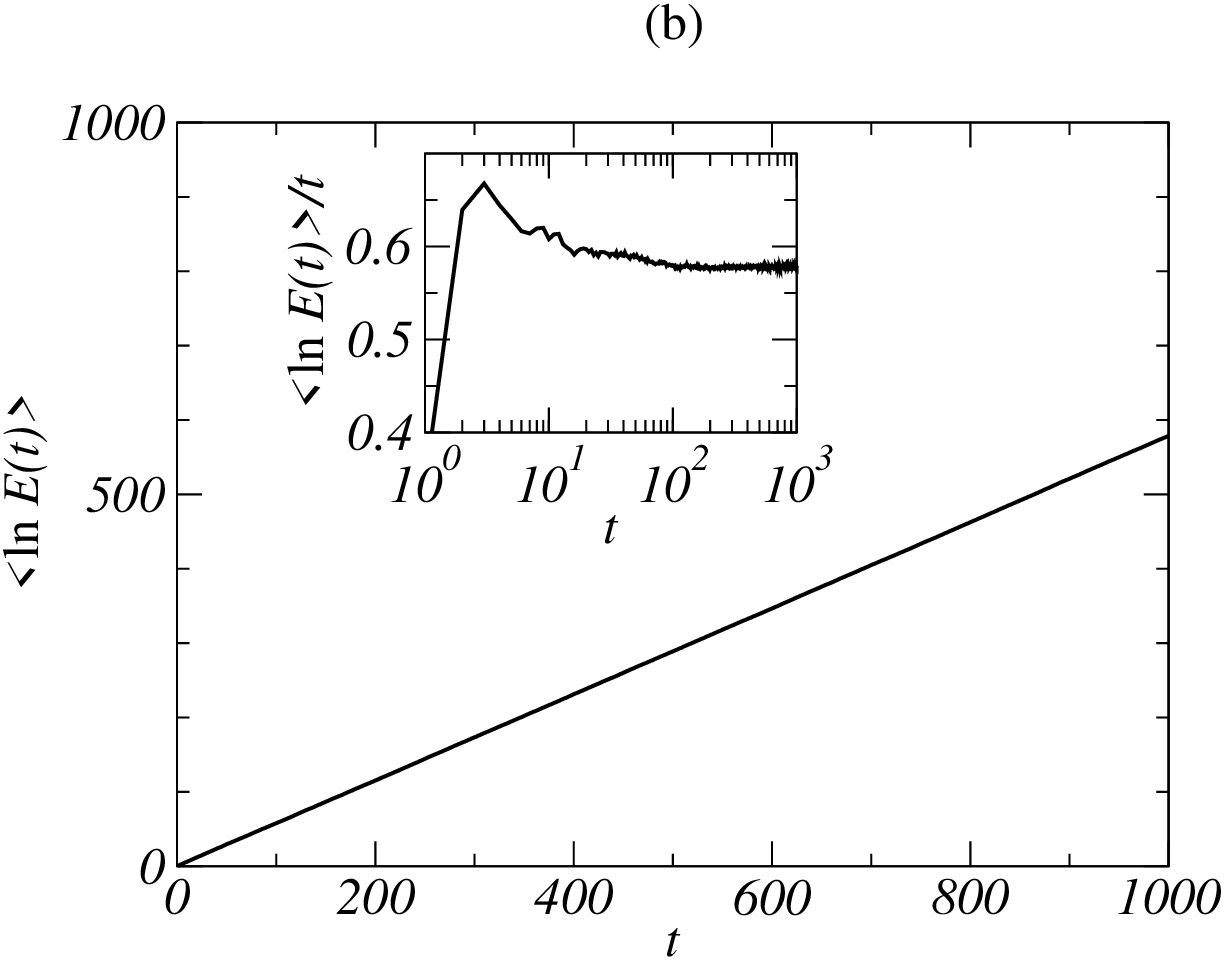}
\hfill

    \caption{\label{fig:OH} Linear oscillator with multiplicative 
noise: Eq.~(\ref{multlin}) is integrated numerically for ${\mathcal D} = 1$ 
with a timestep $\delta t = 10^{-3}$. Ensemble averages are computed 
over $10^4$ realizations. We plot the average 
$\langle \ln E(t) \rangle$ and the ratio $\langle \ln E(t) \rangle/t$ 
(inset) \emph{vs.} time $t$. Fig.~(a): $\omega = 1$, 
$\Lambda(\omega = 1) = \lim_{t \rightarrow \infty} 
\langle \ln E(t) \rangle/t = 0.219(3)$;
Fig.~(b): degenerate case $\omega = 0$,
$\Lambda(\omega = 0) \simeq 0.580(5)$.
Both estimates of the Lyapunov exponent are in good agreement 
with the theoretical prediction \cite{hansel,tessieri}.
}
\end{figure}

  It has also  been  proved   that the quenched
 average of the energy $E$  grows linearly with time, hence
   the   Lyapunov exponent $\Lambda$, defined as
\begin{equation}
 \Lambda = \lim_{ t \rightarrow \infty} \frac{1}{t}
\, \langle \log E  \rangle,
\label{deflambda}
\end{equation}
 is  finite and strictly positive  \cite{hansel,tessieri}.
 The   positivity  of  the   Lyapunov exponent    implies the
 instability of 
 all   moments  at long  times. Note that  the growth rate  $\mu$,
 defined in  Eq.~(\ref{moyenerg}),  
  is  larger than the Lyapunov exponent 
 because of the convexity inequality,  $\log \langle E \rangle
 \ge \langle \log E \rangle$.

 In Figure \ref{fig:OH}(a), we present the   numerical solution
  of Eq.~(\ref{multlin})  averaged over a large number of
 realizations of the noise, where the pulsation is  $\omega = 1$. 
 The algorithm used  to solve 
  this  stochastic differential
 equation with multiplicative noise 
 is inspired from  \cite{mannella} and  explained
 in Appendix \ref{sec:num}.   A  numerical estimate  of the 
 Lyapunov exponent, given in Fig.~\ref{fig:OH}(a),
 agrees very well with the analytic expression of 
 Refs.~\cite{hansel,tessieri}.
 The usual statistical  equipartition of the total energy between   
  kinetic and   potential  contributions  is satisfied:
 $\langle E \rangle   = \omega^2 \, \langle x^2 \rangle = 
 \langle v^2 \rangle $.
  
   In Figure \ref{fig:OH}(b), we show the same quantities for the degenerate
 linear oscillator obtained by taking  $\omega$  equal to 0. This
 degenerate case exhibits the same behavior as the generic case and 
 the Lyapunov exponent  can  be calculated  by  taking  
 the $\omega \to 0$ limit in the formulas  of  Refs.~\cite{hansel,tessieri}.
  We conclude that 
  the instability triggered by the noise  is the dominant
 effect and  that  the presence of the linear restoring force
 $-\omega^2 x$ is irrelevant.

  Hence,  in order  to avoid an exponential 
  increase of the energy
 and the amplitude of the oscillator, it is necessary to 
 go beyond the linear approximation and consider the effect
  of nonlinear restoring forces \cite{degli}.

 \section{Duffing  oscillator with multiplicative noise}
\label{sec:duff}  

  We  now  analyze  the effect of a  nonlinear
  restoring force  in an  oscillator submitted  to  an external
   multiplicative noise.
 In order to preserve the $x \to -x$ symmetry,
 the   nonlinear term has
 to be odd  in the amplitude $x$. In this section, we study the
  particular case of a  cubic nonlinearity
\begin{equation}
   \frac{\textrm{d}^2 }{\textrm{d} t^2}x(t) +(\omega^2 + \xi(t))\, x(t) 
   + \lambda \, x(t)^3  = 0.
 \label{multnonlin}
\end{equation}
The coefficient of the nonlinear term  is   set equal to unity
 by    rescaling the variable
  $x(t)$   to   $ x(t) \,\sqrt{\lambda}$. The random noise 
  $\xi$ is  a  Gaussian white noise of amplitude  ${\mathcal D}$
 as  defined in Eq.~(\ref{defgamma}). 
 The deterministic  nonlinear mechanical system  
 correponding to Eq.~(\ref{multnonlin})
 is known as the Duffing oscillator. We shall prove that  
 the cubic term is relevant and  prevents
  the  average amplitude  
 from growing   exponentially. Instead of
 an exponential behavior, 
  the average energy  of the oscillator 
  as well as the variances of its  position and velocity
  exhibit  a power-law behavior  with time. We shall 
   calculate exactly the associated  scaling  exponents.

    \subsection{The  degenerate  cubic oscillator}
\label{sec:duff:deg}

\begin{figure}[ht]
\hfill
\includegraphics*[width=0.4\textwidth]{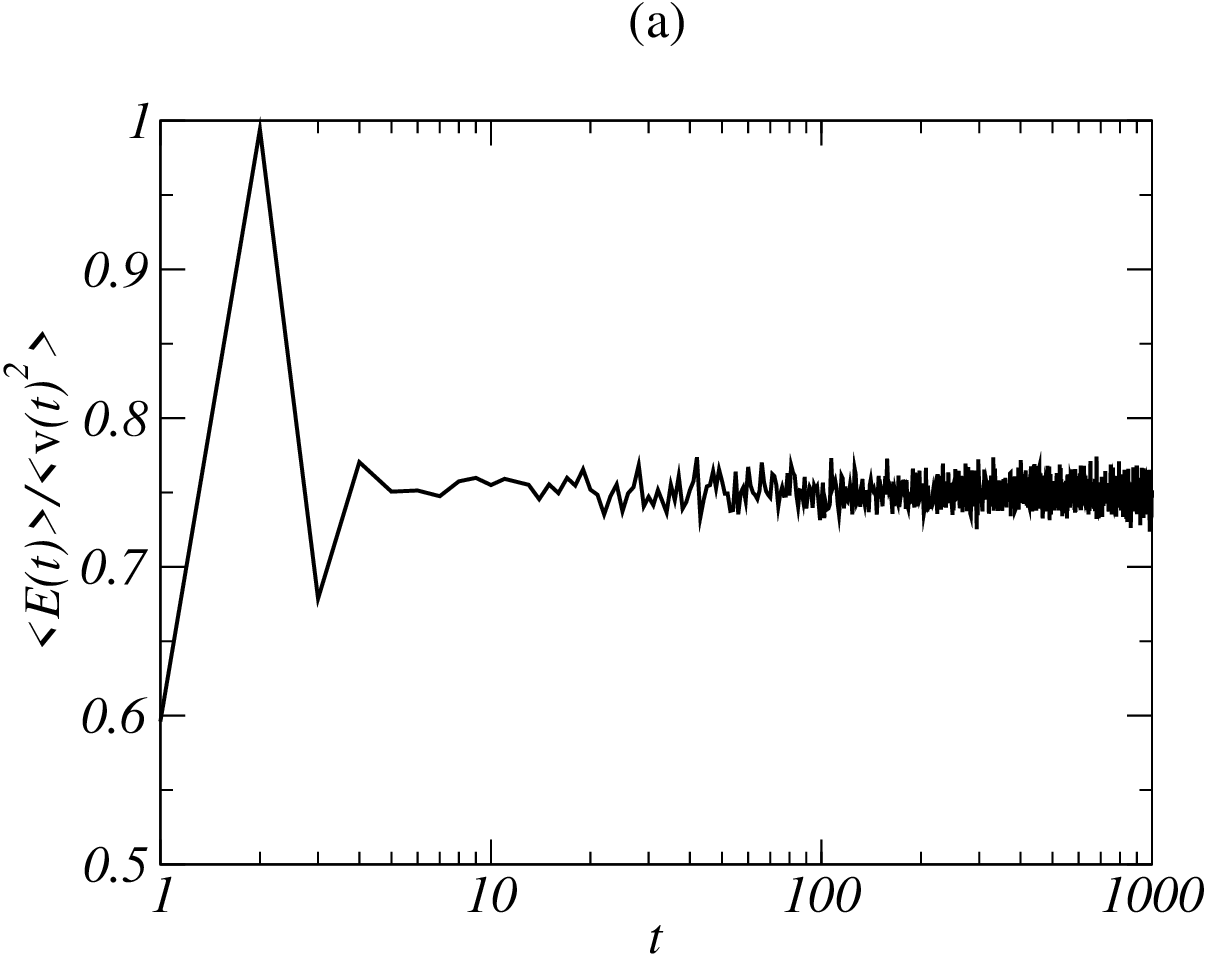}    
\hfill
\includegraphics*[width=0.4\textwidth]{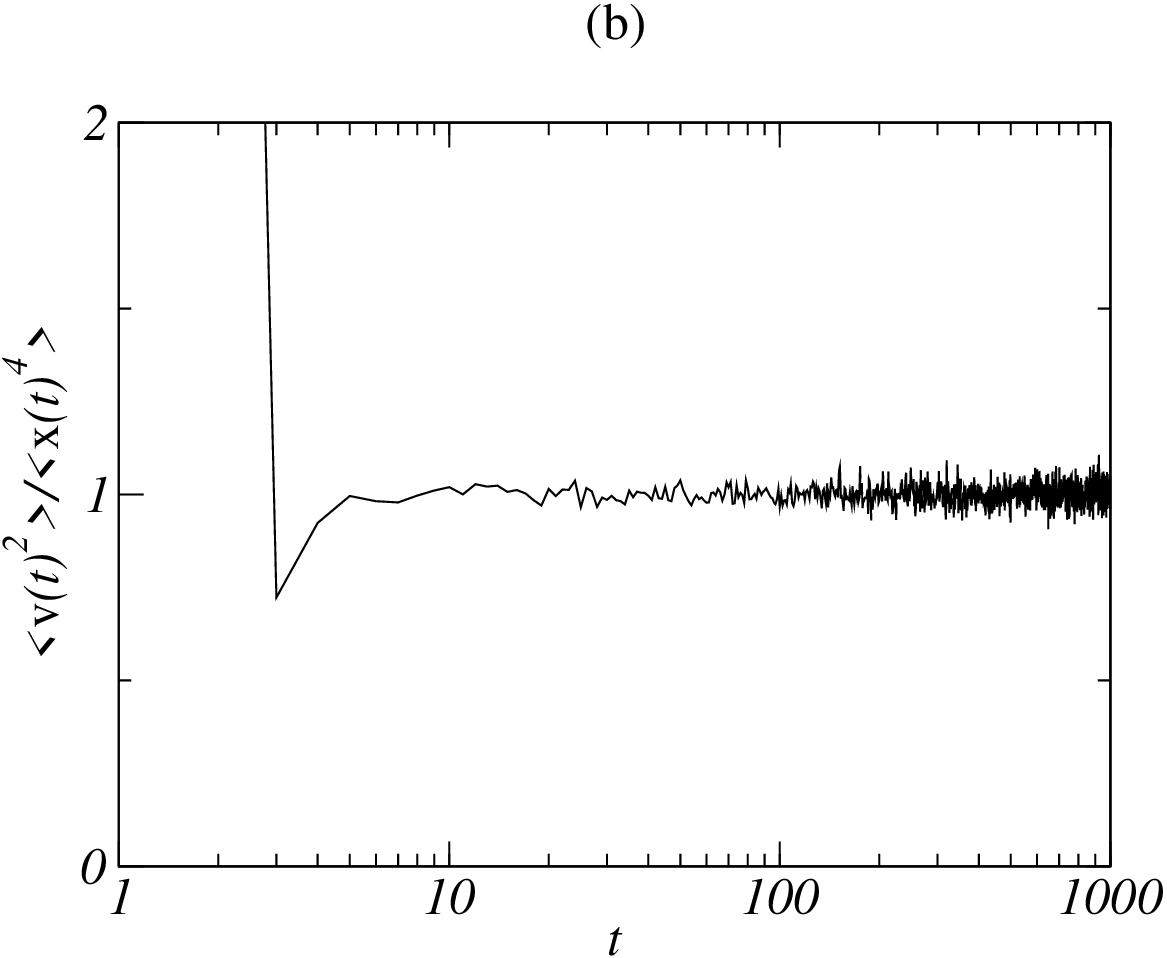}    
\hfill

    \caption{\label{fig:cubic:equi} Cubic oscillator: 
energy equipartition.
 Eq.~(\ref{cubic}) is integrated numerically for ${\mathcal D} = 1$
with a timestep $\delta t = 5 \, 10^{-4}$. Ensemble averages are
computed over $10^4$ realizations.
 The ratios 
$\langle E(t) \rangle / \langle v(t)^2 \rangle$
and $\langle v(t)^2 \rangle / \langle x(t)^4 \rangle$
are approximately constant and respectively equal to $3/4$ and $1$.
}
\end{figure}

  The   linear part  of the restoring
 force, $-\omega^2 x$,  is   negligible in comparison to 
  the  cubic term when   the amplitude
 of the oscillator  is large. 
 In order to study the  long-time  behavior  
 of the oscillator,  we therefore simplify 
   Eq.~(\ref{multnonlin})   to that   of a degenerate
 cubic oscillator:
 \begin{equation}
\frac{\textrm{d}^2 }{\textrm{d} t^2}x(t) + \xi(t) \, x(t) + x(t)^3  = 0.
 \label{cubic}
\end{equation}
  We  first  study   the deterministic
 part of Eq.~(\ref{cubic})  and shall  add the noise term  afterwards
  \cite{roberts}.   
 In one dimension, the  deterministic  cubic  oscillator  is 
 integrable   because  
 the energy $E$  defined as
\begin{equation}
   E =  \frac{1}{2}\dot x^2 + \frac{1}{4} x^4,
\label{cubicE}
\end{equation}
 is a conserved quantity.
 The exact solution of the mechanical  system,
 $\ddot x + x^3 = 0$,
  for a fixed value of  $E$,  is given by 
 \begin{eqnarray}
          x &=&   E^{1/4}  \,
 \frac{ {\rm sn}\left((4E)^{\frac{1}{4}}t;\frac{1}{\sqrt{2}}  \right)}
      { {\rm dn}\left((4E)^{\frac{1}{4}}t;\frac{1}{\sqrt{2}}  \right)}  ,
  \label{solx}  \\                 
     \dot x &=&  (2E)^{1/2} \,
 \frac{ {\rm cn}\left((4E)^{\frac{1}{4}}t;\frac{1}{\sqrt{2}}  \right)}
      { {\rm dn}^2\left((4E)^{\frac{1}{4}}t;\frac{1}{\sqrt{2}}  \right)} ,
\label{solv}
 \end{eqnarray}
 where sn, cn and dn are the classical Jacobi  elliptic functions
 of modulus $ 1/\sqrt{2}$  \cite{abram,byrd}.
 The properties of elliptic functions  that will be
 useful  for our purpose  are recalled in Appendix \ref{sec:elliptic}.
 From  Eqs.~(\ref{solx}) and (\ref{solv}),
 we define the action-angle
 variables of the  cubic oscillator \cite{landau,lichtenberg}.
 The action variable $I$ corresponds to 
  the area  under a  constant
 energy curve in phase space:
 \begin{equation}
 I  = 4 \int_0^{(4E)^{\frac{1}{4}}} \sqrt{ 2E - \frac{x^4}{2}} \, \textrm{d}x
     = 8 \, \left(\int_0^1  \sqrt{ 1 - u^4} \textrm{d}u \right)\, E^{3/4} 
     \propto   E^{3/4} .
 \label{aire}
\end{equation}
 The angle variable $\phi$,  canonically 
conjugate to the action $I$,  
 is equal to  $\omega(E) t$ (but for an unimportant  additive  constant) 
  where $\omega(E)$ is the frequency
 corresponding to the  energy   $E$. From  Eqs.~(\ref{solx}) and 
 (\ref{solv}),  the phase  is  identified  as
\begin{equation}
  \phi = (4E)^{\frac{1}{4}} \, t  .
 \label{angle}
\end{equation}
 This phase variable is  defined  modulo the common  real  period
 $4K$ of the elliptic functions ${\rm sn},{\rm cn},{\rm dn}$ 
 that appear in Eqs.~(\ref{solx}) and (\ref{solv}).
 The quarter of the period  is given by
\begin{equation}
     K = K(\frac{1}{\sqrt{2}}) = \sqrt{2}
  {\int_0^1 \frac{\textrm{d} u} {\sqrt{ 1 - u^4 }}}  \simeq  1.854 \, .
\label{periode}
\end{equation}
 In terms of the energy-angle  coordinates  $(E, \phi)$,
 the original variables  $(x ,\dot x)$ read as follows: 
 \begin{equation}
          x =   E^{1/4}  {\rm sd}(\phi;{1}/{\sqrt{2}})   
    \,\,\,      \hbox{ and  } \,\,\,                 
     \dot x =  (2E)^{1/2}
 \frac{ {\rm cn}(\phi;{1}/{\sqrt{2}})}
      { {\rm dn}^2(\phi;{1}/{\sqrt{2}}) } , 
\label{solxv2}
 \end{equation}
  where we have introduced the function sd = sn/dn.
  Hence,  for the free system  (without noise) 
 the  second-order dynamical equation, $\ddot x + x^3 = 0$,
 is equivalent to  the  following  two 
   first-order equations,  the first one representing 
  energy  conservation:
 \begin{eqnarray}
 \dot E  &=&  0  \, ,\nonumber \\
 \dot \phi &=&  (4E)^{\frac{1}{4}}   \, .
\label{enphi}
 \end{eqnarray}

The presence of external  noise spoils the
 integrability of the  dynamical system  (\ref{cubic})
  and causes $E$  to grow with time  
  by continuously  injecting energy into the system.
 We  now take into account the external  noise
  and  rewrite   the system  (\ref{enphi})
  in  $(E, \phi)$ coordinates.   
 The  stochastic   evolution 
  equation  for   the energy is given by
  \begin{equation}
 \dot E = x \dot x \xi(t) = \sqrt{2} \, E^{3/4} \,
  \frac{ {\rm sn}(\phi;{1}/{\sqrt{2}}) \, {\rm cn}(\phi;{1}/{\sqrt{2}})   }
      { {\rm dn}^3(\phi;{1}/{\sqrt{2}})} \, \xi(t)  \,.
\label{evolE} 
\end{equation}
 Inverting the relation (\ref{solxv2})  
  between $\phi$ and $x$,  
  \begin{equation}
  \phi = {\rm sd}^{-1}\left(\frac{x}{E^{1/4}}, \frac{1}{\sqrt{2}}\right)\,  , 
\label{cubicangle}
   \end{equation}
 and using Eqs.~(\ref{solxv2})  and (\ref{evolE}), we obtain 
  the stochastic  evolution of  the phase variable:
 \begin{equation}
 \dot\phi = (4E)^{\frac{1}{4}} - \frac{ {\rm sd}^2(\phi;{1}/{\sqrt{2}})}
{ 2\sqrt{2} E^{1/4}} \xi(t) \,   .
 \label{evolangle}
\end{equation}
   With the help of the  auxiliary  variable $\Omega$  defined as 
 \begin{equation}
  \Omega  = 2\sqrt{2} E^{1/4}  \, ,
 \label{action1}
\end{equation}
 we  derive    a  compact  and symmetric form for 
  the two stochastic  evolution  equations 
  (\ref{evolE}) and (\ref{evolangle}):
   \begin{eqnarray}
     \dot \Omega  &=&  \frac{ {\rm sn}(\phi){\rm cn}(\phi)   }
      { {\rm dn}^3(\phi)} \xi(t)    ,
   \label{evolOmega} \\
  \dot\phi  &=& \frac{\Omega}{2} - \frac{ {\rm sd}^2(\phi)}{\Omega}  \xi(t)   .
    \label{evolphi}
   \end{eqnarray}
  where the elliptic modulus $1/\sqrt{2}$
 common to all the elliptic functions  has been omitted.
 We  emphasize that the coupled equations (\ref{evolOmega})
 and (\ref{evolphi}) are mathematically
 equivalent to the initial system and 
  have been derived without any approximation. Moreover,
 the  nature of the parametric perturbation  has played no role
 in the derivation: the function  $\xi(t)$ can be a  deterministic
 or a  stochastic function with arbitrary statistical properties.
 
 \hfill\break

\begin{figure}[th]
\hfill
\includegraphics*[width=0.4\textwidth]{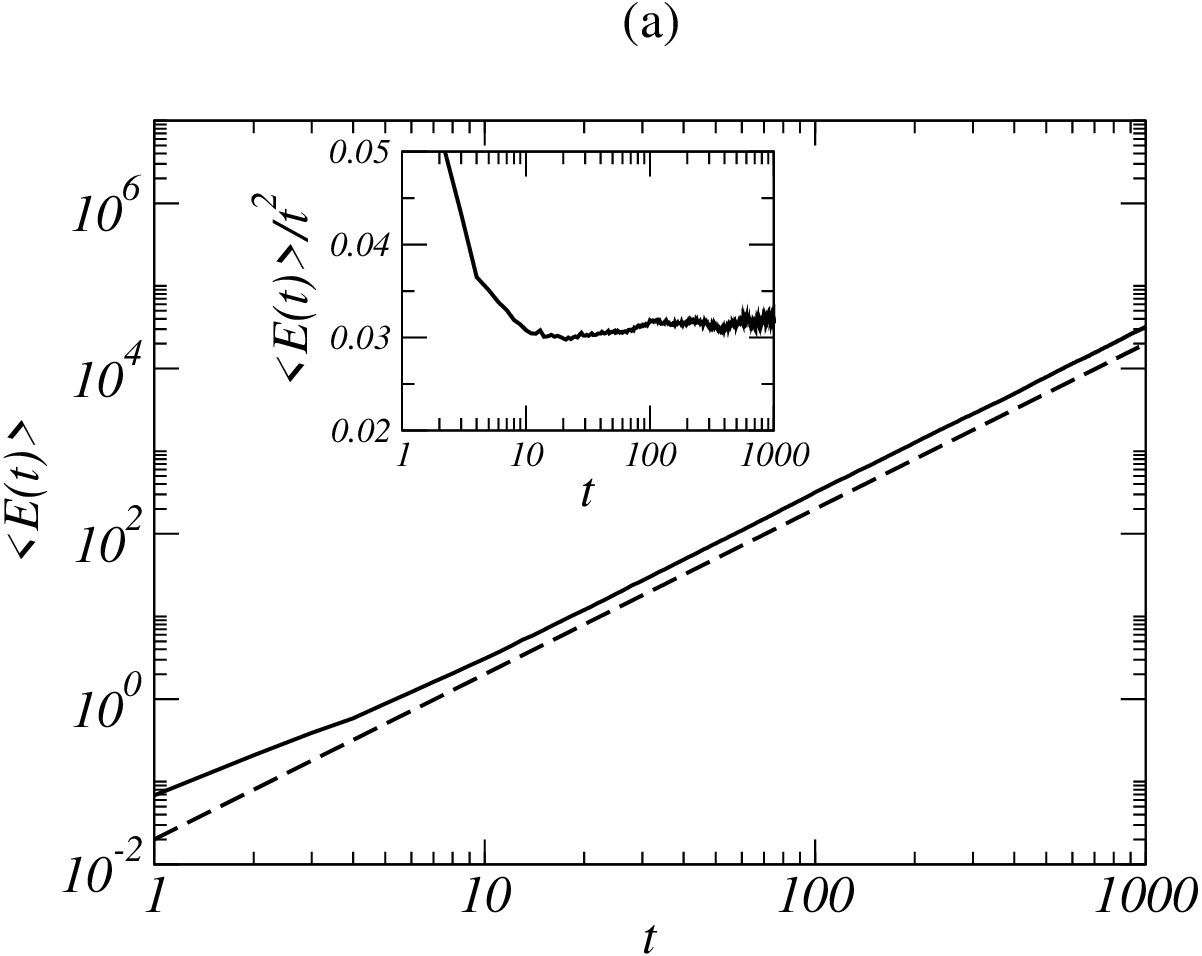}    
\hfill
\includegraphics*[width=0.4\textwidth]{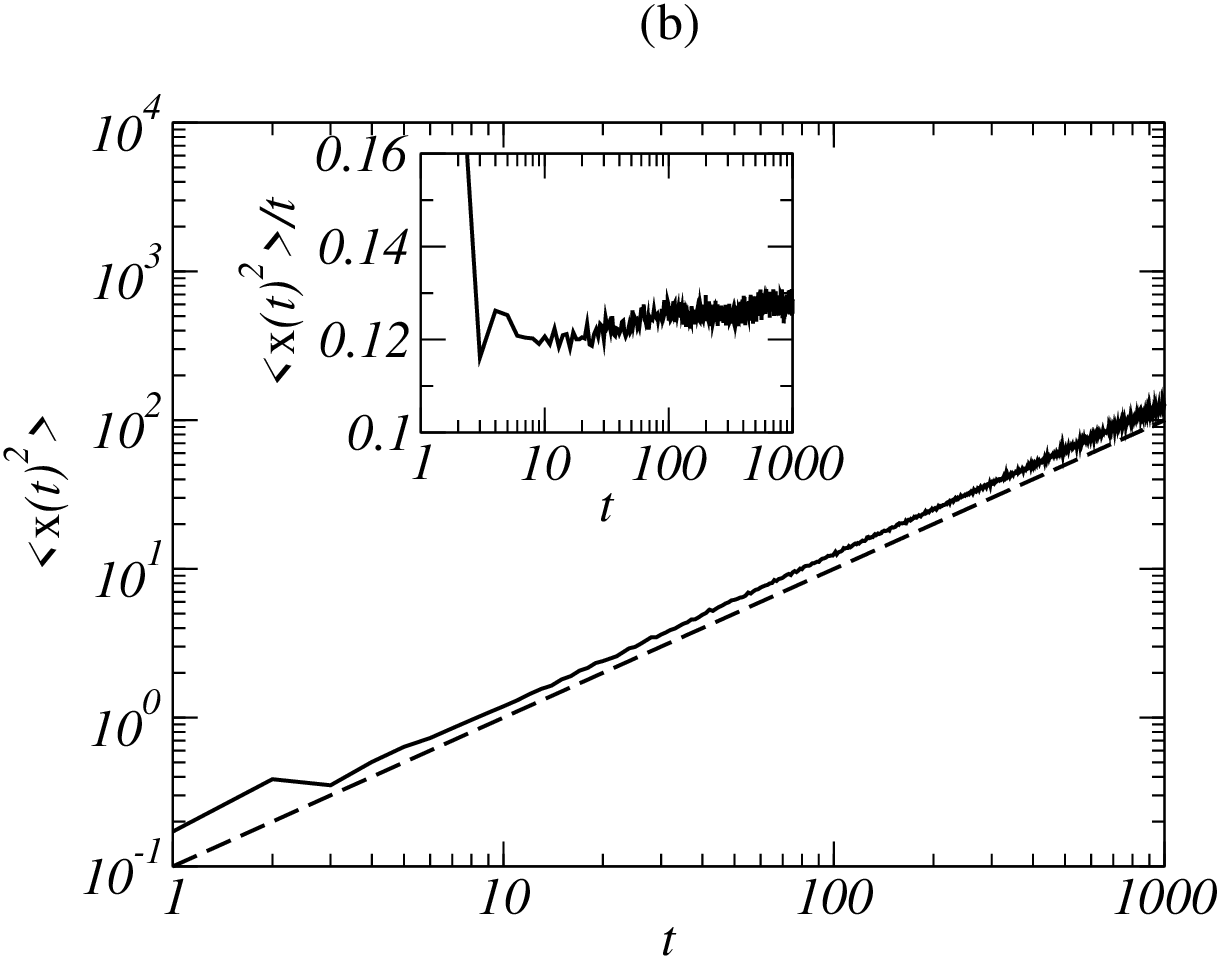}    
\hfill
\vspace*{0.1cm}
\hfill
\includegraphics*[width=0.4\textwidth]{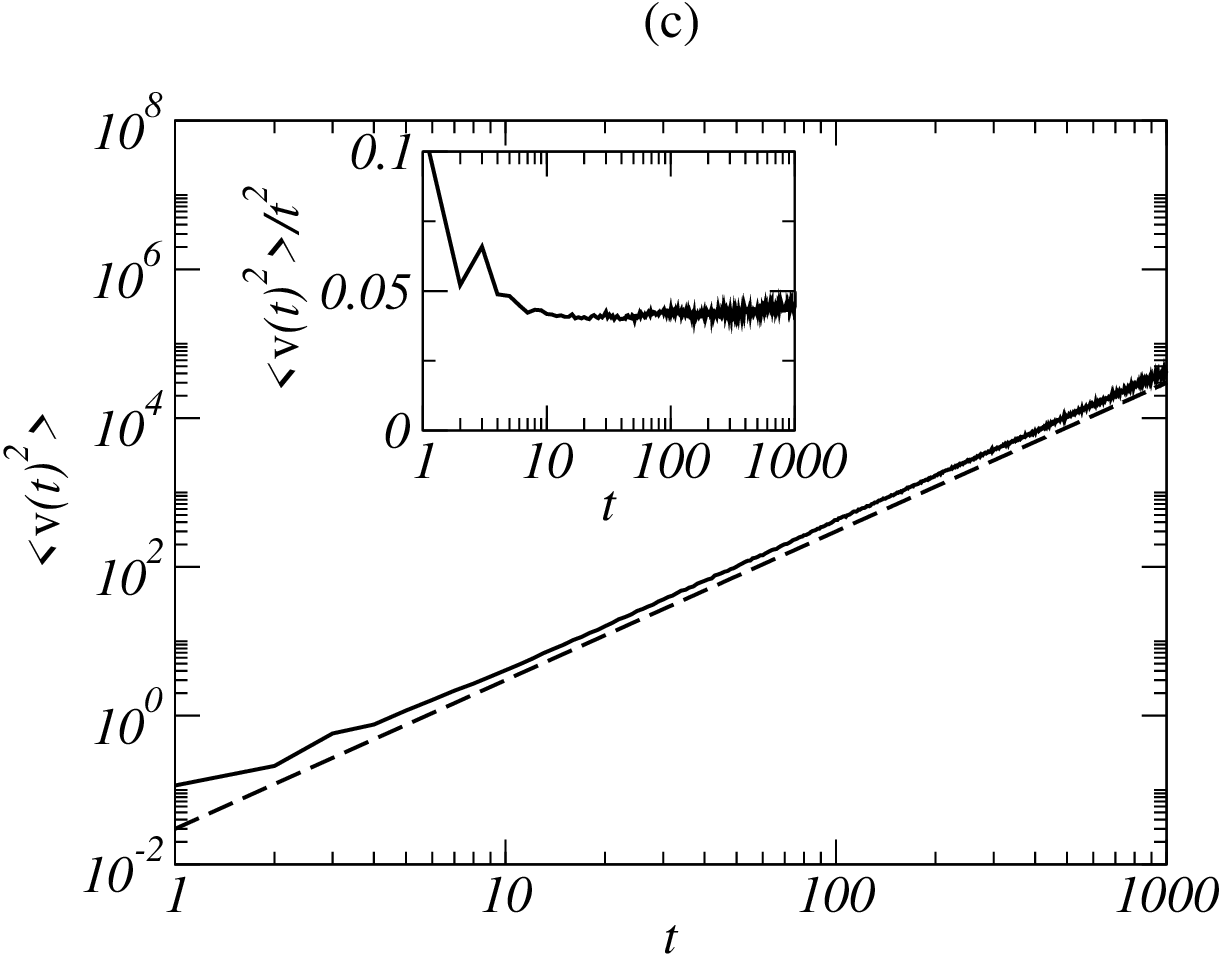}    
\hfill

    \caption{\label{fig:cubic:scaling} Cubic oscillator: 
numerical data, obtained for the same runs as in 
Fig.~\ref{fig:cubic:equi}, is represented
by solid lines, and compared to the predicted power-law behaviors given
by Eqs.~(\ref{scalings})  (dashed lines). We plot the time-dependence
of the following averages: (a) $\langle E(t) \rangle$, (b)
$\langle x(t)^2 \rangle$ and (c) $\langle v(t)^2 \rangle$.
Data plotted in the insets yield the following estimates of the 
diffusion constants $D_E = 0.031(1)$, $D_x = 0.125(5)$, $D_v = 0.043(3)$,
in excellent agreement with theoretical estimates Eqs.~(\ref{Diffenerg}), 
(\ref{diffspeed}) and   (\ref{diffposit}).
}
\end{figure}

 We now  perform  a precise analysis 
  of  the long-time behavior of the cubic  oscillator 
 driven by a  multiplicative Gaussian  white noise.
 From a heuristic  point of view,  
 we  observe   from  Eq.~(\ref{evolOmega})
  that   $\Omega$   undergoes
 a diffusion process and   should  scale  typically
 as  $t^{1/2}$. 
  We  also  notice    from  Eq.~(\ref{evolphi})
 that,  as $\Omega$ grows,   the phase  $\phi$ varies 
  more and more rapidly with time. Hence
  the phase  $\phi$  is a fast variable and it is natural  to 
 average the dynamics  over its  rapid variations.  
  This averaging process  leads to  some  remarkable
 and  general  identities  between  different physical quantitites.
  Thus,  we obtain
  the  average  of  $ \dot x^2$  from  Eq.~(\ref{solxv2}),
  \begin{equation}
   \langle \dot x^2   \rangle =  2 \, \overline {\frac{ {\rm cn^2}(\phi)}
      { {\rm dn}^4(\phi) }  }\,  \langle  E   \rangle  ,
\label{equip1}
\end{equation}
where the overline indicates the mean-value over a period $4K$.
 The  coefficient  in Eq.~(\ref{equip1})
 can be calculated explicitly:
 \begin{equation}
 2 \, \overline {\frac{ {\rm cn^2}(\phi)}
      { {\rm dn}^4(\phi) }     }
 = \frac{4}{3}  \, ; 
\end{equation}
  the proof of this  identity  (and of some  forthcoming
 similar identities)  is  presented  in Appendix \ref{sec:elliptic}.  
  Equation~(\ref{equip1}) is nothing but a
  statistical   {\it equipartition } relation between
 the mean values  of the total  energy and  kinetic energy:
\begin{equation}
 \langle  E   \rangle = \frac{3}{4}  \langle \dot x^2   \rangle .
\label{equipartition}
\end{equation}
  Such an identity provides a  generalization of 
  the usual equipartition theorem valid
 for quadratic degrees of freedom.
  Another statistical equality follows  from  Eq.~(\ref{equipartition}) and
   the definition (\ref{cubicE})  of the energy $E:$
 \begin{equation}
 \langle \dot x^2   \rangle  = \langle  x^4   \rangle .
\label{equipxv}
\end{equation}

 We emphasize that these equalities  are `universal' in the sense that  they
 are independent of  the form of the noise  we consider.
 The derivation of these equipartition relations  relies  only
 on the hypothesis  that  the distribution of the angle
 $\phi$ is   uniform over the interval $[0, 4K]$  when $ t \to \infty $.
 In particular,    identities  (\ref{equipartition}) and (\ref{equipxv})
 are  valid for multiplicative as well as for additive noise
 (see Appendix \ref{sec:add}).
 Our numerical results  (Fig.~\ref{fig:cubic:equi}) 
 confirm  these relations and that
  the averaging procedure   is   therefore  valid.

\begin{figure}[ht]
\hfill
\includegraphics*[width=0.4\textwidth]{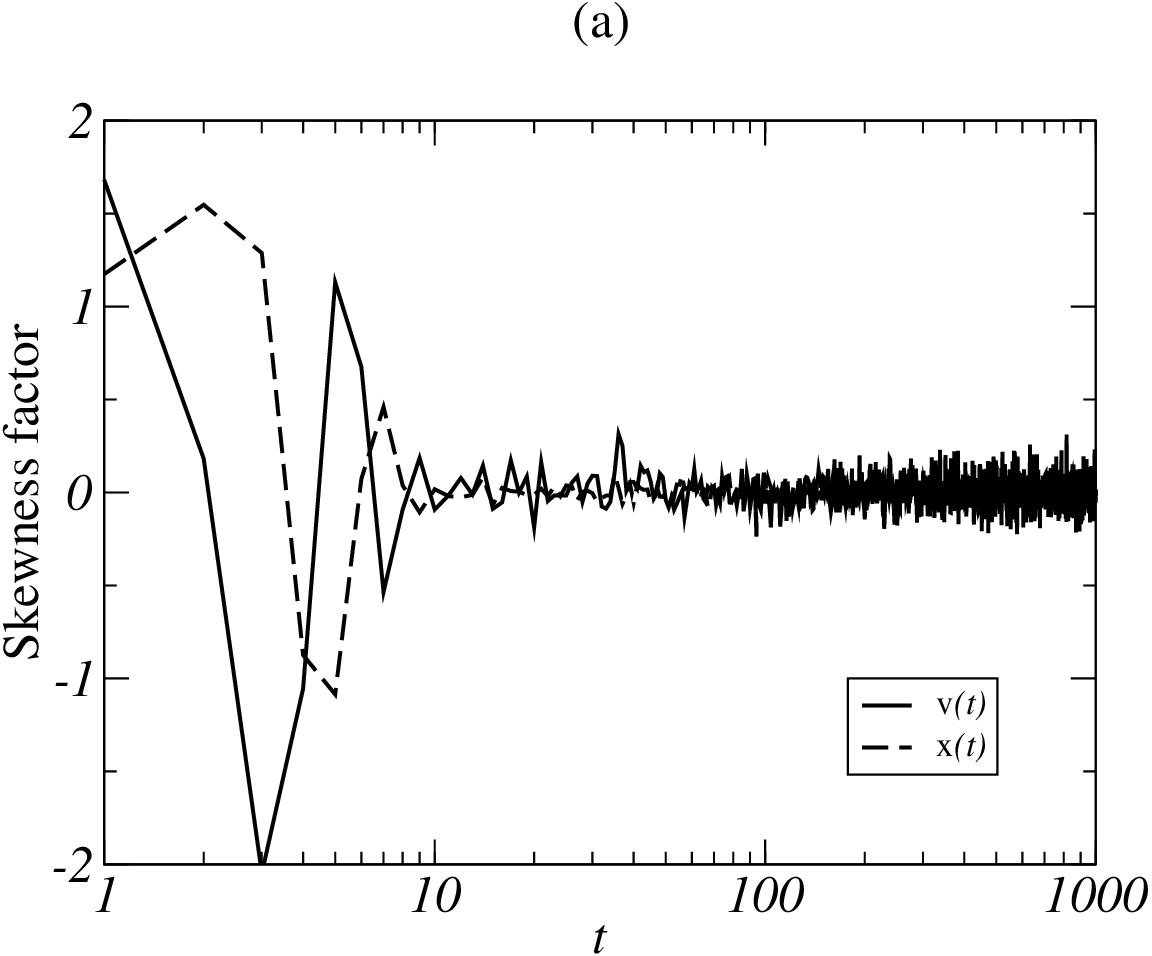}    
\hfill
\includegraphics*[width=0.4\textwidth]{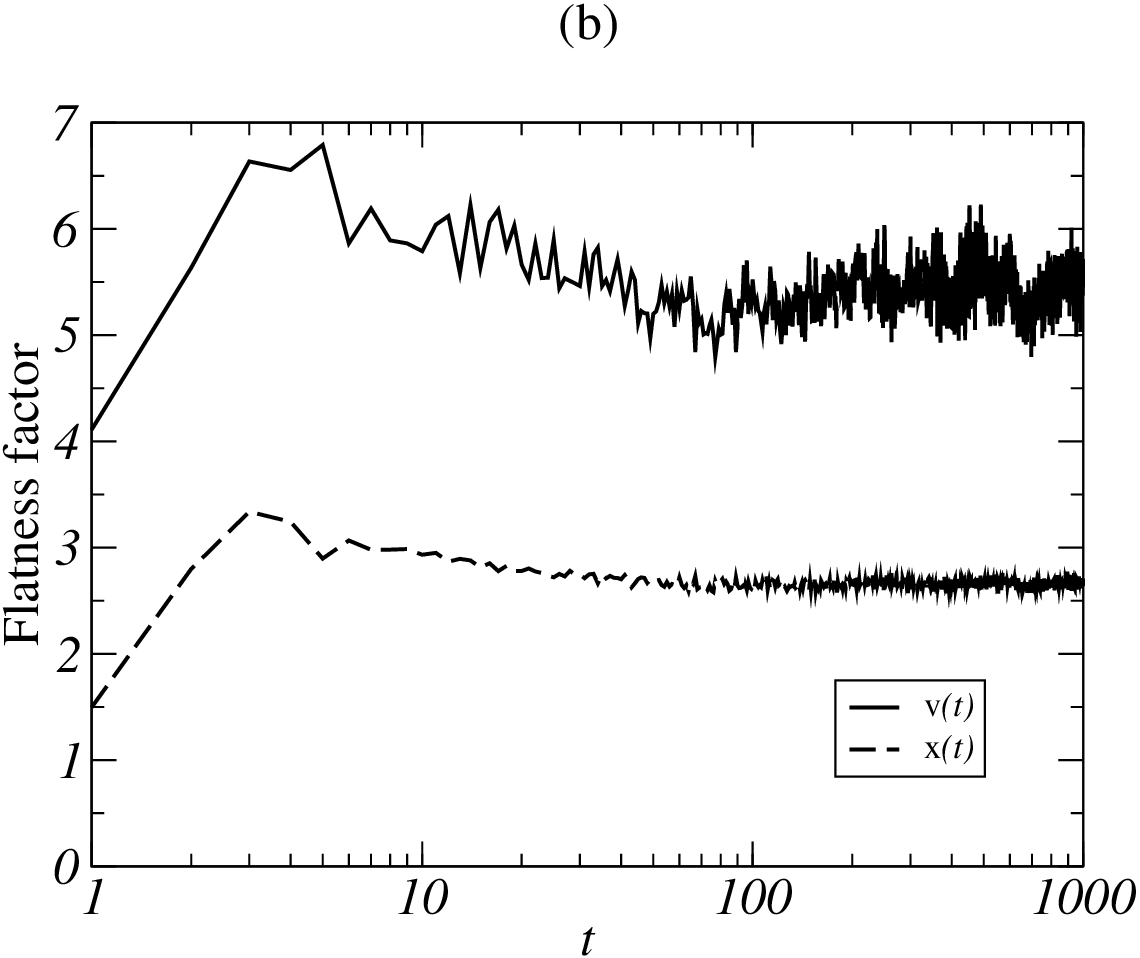}    
\hfill
\vspace*{1.0cm}

    \caption{\label{fig:cubic:Gauss} Cubic oscillator. Fig.~(a):
skewness factors of the position $x(t)$ (dashed line)
and velocity $v(t)$ (solid line). Fig.~(b):
flatness factors of the position $x(t)$ (dashed line)
and velocity $v(t)$ (solid line).
Numerical data is obtained from the same runs as in 
Fig.~\ref{fig:cubic:equi}.
}
\end{figure}

 The same averaging procedure allows us to derive  
  a closed  equation  for  the stochastic
 evolution  of   the slow variable  $\Omega.$ 
  We start by  writing   the Fokker-Planck equation 
  governing  the evolution of the P.D.F. $P_t(\Omega, \phi)$
  associated  with the system (\ref{evolOmega},\ref{evolphi}).
 Since the noise term appears as a multiplicative factor, one  must
 be cautious about  the convention used  to define stochastic calculus.
  Here as well as in the following, we shall use Stratonovich rules because
 they are  obtained naturally  when white noise is considered
 as a limit of colored noise with vanishing correlation time 
  \cite{vankampen,risken}.
 The   Fokker-Planck equation  corresponding
 to Eqs.~(\ref{evolOmega}) and (\ref{evolphi}) reads:
 \begin{eqnarray}
  \partial_t P  &=&  
 - \frac{\Omega}{2} \partial_{\phi}P  \nonumber \\
 &+&  \frac{{\mathcal D}}{2}
\Bigg( \partial_{\phi}\left( \frac{g(\phi)}{\Omega}  \partial_{\phi}
  \frac{g(\phi)}{\Omega} P \right)
 - \partial_{\phi}\left(  \frac{g(\phi)}{\Omega} 
   \partial_{\Omega}P   f(\phi)      \right)  
- \partial_{\Omega}\left( f(\phi) 
 \partial_{\phi} \frac{g(\phi)}{\Omega}P \right)  
+ \partial_{\Omega}\left( f(\phi) \partial_{\Omega}f(\phi)P    \right)  
\Bigg)    \, ,
\label{FPomeg}
 \end{eqnarray}
where we have defined:
 \begin{equation}
  f(\phi)  =  \frac{ {\rm sn}(\phi){\rm cn}(\phi)   }
      { {\rm dn}^3(\phi)} \,\,\, \hbox { and }  \,\,\, 
  g(\phi) = {\rm sd}^2(\phi)  \,\, .
\end{equation}  

 The Fokker-Planck equation (\ref{FPomeg}) written in the
  variables $(\Omega,\phi)$ is exact because we
 study the case of a Gaussian white noise. In order to
 pursue our calculations, we assume that   $P_t(\Omega, \phi)$
  becomes  independent of  $\phi$  when $t \to \infty$, {\it i.e.}, 
 that  the  probability measure  for $\phi$ is  uniform over  
   the   interval $[0,4K]$. 
   We now average the Fokker-Planck
 equation (\ref{FPomeg}) over the angular variable.  We shall use the fact that
 the average of the  derivative of any function is zero:
  \begin{equation}
 \overline{ \partial_{\phi}(\ldots)} = 0  \, .
\end{equation}  
    This implies  that:
\begin{equation}
 \overline{ \partial_{\phi}\left(g(\phi) \partial_{\phi}g(\phi) \right) }
 = 0  \,\,\, \hbox{ and } \,\,\, \overline{ f(\phi) } = 0
  \,\,\, \hbox{  because  } \,\,\, f(\phi) = \frac{1}{2}\partial_{\phi}g(\phi) 
 \, .
\label{idmoy}
\end{equation}  
  Using these properties, and in particular the last identity in
 (\ref{idmoy}), we  derive  the phase-averaged Fokker-Planck 
 equation:
 \begin{equation}
   \partial_t \tilde{P}  =  \frac{{\mathcal M }{\mathcal D} }{2} 
\left(  \partial_{\Omega}^2 \tilde{P} - 2 \partial_{\Omega}
      \frac{\tilde{P}}{\Omega}  \right)  \, ,
\label{FPmoy}
\end{equation}  
 where  $\tilde{P}_t(\Omega)$ is  now a function of $\Omega$ and $t$ only,
 and where ${\mathcal M }$ is given by  
 \begin{equation}
{\mathcal M } = \overline{f^2(\phi)} =    \frac{1}{K} \int_0 ^K
  \frac{ {\rm sn^2}(\phi){\rm cn^2}(\phi)   }
      { {\rm dn}^6(\phi)} \, \textrm{d}\phi =  
    2 \,  \frac{\int_0^1 u^2\sqrt{ 1 - u^4 } \textrm{d}u}
  {\int_0^1 \frac{\textrm{d}u} {\sqrt{ 1 - u^4 }}} 
 =  \frac{96\,\pi^2}{15 \, \left( \Gamma({\frac{1}{4}}) \right) ^4} 
\simeq 0.3655...
\label{defM}
\end{equation}  
 The  third equality is obtained 
  by setting  $u = {\rm sd}(\phi)/\sqrt{2}$;
 the expression in terms of the Euler $\Gamma$
 function can be found in \cite{abram}.
  The averaged Fokker-Planck equation  (\ref{FPmoy})
  corresponds  to  the following  effective
 Langevin dynamics for the variable $\Omega$,
\begin{equation}
   \dot\Omega = \frac{{\mathcal M }{\mathcal D}}{\Omega}  + \Xi(t)  \, ,
\label{Langeff} 
\end{equation}  
where the effective noise $\Xi(t)$  satisfies the relation 
 \begin{equation}
\langle  \Xi(t) \Xi(t') \rangle = {\mathcal M } {\mathcal D} \delta( t  - t').
  \label{bruiteffect}
 \end{equation}
We conclude from Eq.~(\ref{Langeff})  that 
 $\Omega$ exhibits a diffusive behavior.  
  The dynamics of all moments of $\Omega$ can be deduced from
  Eq.~(\ref{FPmoy}), as for example:
\begin{eqnarray}
  \partial_t \langle \Omega^2 \rangle &=&  3 \,  {\mathcal M } {\mathcal D}\, ,
\label{varOm} \\
  \partial_t \langle \Omega^4 \rangle &=& 10  \,   {\mathcal M } {\mathcal D}
  \, \langle \Omega^2 \rangle  \, . \label{tetraOm} 
\end{eqnarray}
 From Eq.~(\ref{varOm}), we calculate  the  effective diffusion constant
 $D_\Omega$ of  $\Omega$:
 \begin{equation}
     D_\Omega  =  3 \,  {\mathcal M } {\mathcal D} \simeq  1.096  \,  
{\mathcal D} \, ,
\label{diffOm}
 \end{equation}
where $\langle \Omega^2 \rangle  \simeq D_\Omega \, t$ when 
$t\rightarrow\infty$.
  We conclude from Eq.~(\ref{tetraOm}) that
  $\Omega$ has a non-trivial flatness:
 \begin{equation}
    \frac{\langle \Omega^4 \rangle} {  \langle \Omega^2 \rangle^2 }
 = \frac{5}{3}  \, .
\label{flatOm}
 \end{equation}
  This  last equality  shows that   although the variable
 $\Omega$   diffuses  as  $t^{1/2}$,  it  is  not  Gaussian  
(and hence it is   not a   Brownian variable).

  From this statistical information about  $\Omega$, we now  derive 
  the scaling  with time of the average of $E$, $x^2$, $v^2$
 and  calculate the numerical  prefactors
 (generalized diffusion constants).
 From  Eqs.~(\ref{action1}),
 (\ref{diffOm}) and (\ref{flatOm}), we deduce that: 
 \begin{equation}
 \langle E  \rangle = \frac{\langle \Omega^4 \rangle}{64}
  = \frac{5}{3} \frac{\langle \Omega^2 \rangle^2}{64}
 = \frac{15}{64} {\mathcal M}^2  {\mathcal D}^2  \, t^2 
  =  \frac{48 \, \pi^4}{5 \, \left( \Gamma({\frac{1}{4}}) \right) ^8}
      {\mathcal D}^2 \, t^2 \,
      \simeq 0.0313 \, {\mathcal D}^2 \, t^2 \, .
 \label{Diffenerg}
\end{equation}
 Using  the equipartition relations (\ref{equipartition})
  and (\ref{equipxv}), we find  that:
\begin{eqnarray}
     \langle \dot x^2   \rangle  &=& \frac{4}{3} \, \langle E  \rangle 
     =   \frac{64 \, \pi^4}{5 \left( \Gamma({\frac{1}{4}}) \right) ^8}   
{\mathcal D}^2  t^2  \,
 \simeq  0.0417 \,  {\mathcal D}^2  t^2    \label{diffspeed} \, ,\\
  \langle  x^2  \rangle  &=&\frac{\overline{{\rm sd}^2(\phi)}}{8} \,
 \langle \Omega^2  \rangle 
 =  \frac{15}{16} {\mathcal M}^2 {\mathcal D} \, t  \, 
 =  \frac{192 \, \pi^4}{5 \left( \Gamma({\frac{1}{4}}) \right) ^8} 
  {\mathcal D} \, t 
 \simeq    0.125  \,  {\mathcal D} \, t   \, .
 \label{diffposit}
\end{eqnarray}
    The prefactor in the mean value of the energy,  Eq.~(\ref{Diffenerg}), 
  agrees with that  of  \cite{lind2}.
   We have verified numerically  the  power-laws 
  predicted by Eqs.~(\ref{Diffenerg}), (\ref{diffspeed}) and 
  (\ref{diffposit}) and determined  the prefactors.
  The results are displayed in Fig.~\ref{fig:cubic:scaling}.
 The numerical values for the  exponents and the  prefactors
  are in perfect agreement with the analytical calculations:
 $ D_E = 0.0313 \, {\mathcal D}^2$, $D_x = 0.125  \,  {\mathcal D}$
 and    $D_v = 0.0417 \,  {\mathcal D}^2$.

   The  averaged  distribution
 fonction ${\tilde P}_t(\Omega)$ can  be  calculated  because
 Eq.~(\ref{FPmoy}) is exactly solvable due to its  invariance 
 under rescalings $ t \rightarrow \lambda^2  t $ , 
  $ \Omega  \rightarrow \lambda \Omega \, , $  $ \lambda $  being  an arbitrary
 real number (this invariance is the same as that of the heat equation).
  Equation  (\ref{FPmoy}) is solved  by using the self-similar  Ansatz:
 $$ {\tilde P}_t(\Omega) = 
        \frac{1}{\sqrt{t}} \, \Pi( \frac{ \Omega}{\sqrt{t}}) \, ,$$
  and  the P.D.F of  $ \Omega $ is found to be 
\begin{equation}
   {\tilde P}_t(\Omega) =
 \sqrt{ \frac{2}{\pi}} \, \Omega^2  \, 
 \frac{ {\rm e}^{ -  \Omega^2/(2{\mathcal D} {\mathcal M} t) }  }
{ ({\mathcal D} {\mathcal M} t )^{\frac{3}{2}}    } \, .
\label{pdfOm}
\end{equation}
 This distribution function  allows us to calculate the P.D.F.  in
 the $(x,\dot x)$ variables in the time-asymptotic regime.
 In particular, the 
  skewness and the flatness factors of the position and  of the velocity
 can  be calculated analytically.
  Since both variables $x$ and
 $\dot x$ are parity symmetric,  their skewness
  vanishes. The   flatness is given by 
 \begin{eqnarray}
  \frac{\langle x^4 \rangle} {  \langle x^2 \rangle^2 }
  &=& \frac{4}{3  \left(\overline{{\rm sd}^2(\phi)}\right)^2  }
\, \frac{\langle \Omega^4 \rangle} {  \langle \Omega^2 \rangle^2 }
 = \frac{16}{ 45  {\mathcal M}^2}
 \simeq 2.66       \, ,   \\ 
\frac{\langle v^4 \rangle} {  \langle v^2 \rangle^2 }  &=&
  \frac{9}{4} \, \overline{ \frac{{\rm cn}^4(\phi)}{{\rm dn}^8(\phi)}  }
 \, \frac{\langle \Omega^8 \rangle} {  \langle \Omega^4 \rangle^2 }
  = \frac{27}{5} = 5.4     \, .
 \end{eqnarray}
 These  values are in excellent agreement   with the numerical
  computations shown in   Fig.~\ref{fig:cubic:Gauss}.
 We notice  that the variables
 $x$ and $\dot{x}$ are   non-Gaussian because
   their flatness  differs from the Gaussian  value 3.
  
To summarize,   the  following scalings have been derived:
\begin{eqnarray}
          E      &\sim&  t^2  ,  \nonumber  \\ 
          x      &\sim&  t^{\frac{1}{2}}  , \nonumber    \\ 
         \dot x  &\sim&  t .
\label{scalings}
\end{eqnarray}
  In particular, we  conclude  that  the amplitude  $x$ of
a degenerate cubic oscillator does not grow
 exponentially with time but has a normal diffusive behavior. 
\hfill\break

 \subsection{The Duffing  oscillator}
\label{sec:duff:osc}  

 We now study the general  case of a non-zero
 pulsation $\omega$:
\begin{equation}
   \frac{\textrm{d}^2 }{\textrm{d} t^2}x(t) +(\omega^2 + \xi(t))\, x(t) 
   + x(t)^3  = 0.
 \label{multnonlin2}
\end{equation}
Here the coefficient of the nonlinear term  is taken to be unity and 
the random noise is Gaussian and white  as  defined in 
Eq.~(\ref{defgamma}). 

  The results of sections~\ref{sec:linear}  and  \ref{sec:duff:deg}
 show two regimes:  starting from a small initial condition, 
 the amplitude of the oscillator grows exponentially with time
 until  $ x \sim \omega$ where the  linear and  nonlinear
 terms  are of the same order and then  the amplitude grows 
 as the square-root of time according to Eq.~(\ref{scalings}).
  Because the  deterministic system corresponding to Eq.~(\ref{multnonlin2}) 
 is  integrable with the help of elliptic functions, 
 this crossover from exponential to  algebraic can be derived in a  
 quantitative manner.

\begin{figure}[ht]
\includegraphics*[width=0.4\textwidth]{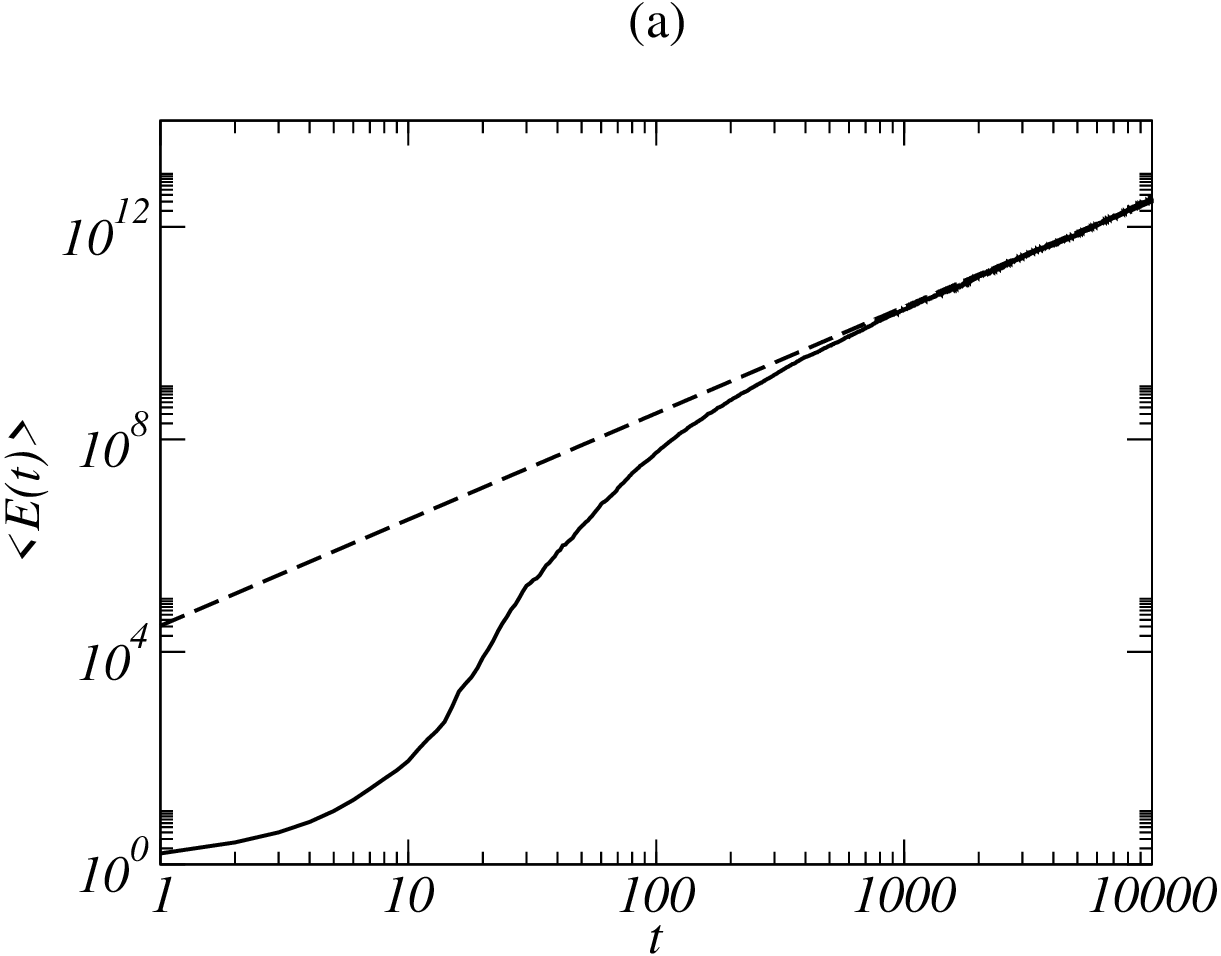}
\hfill
\includegraphics*[width=0.4\textwidth]{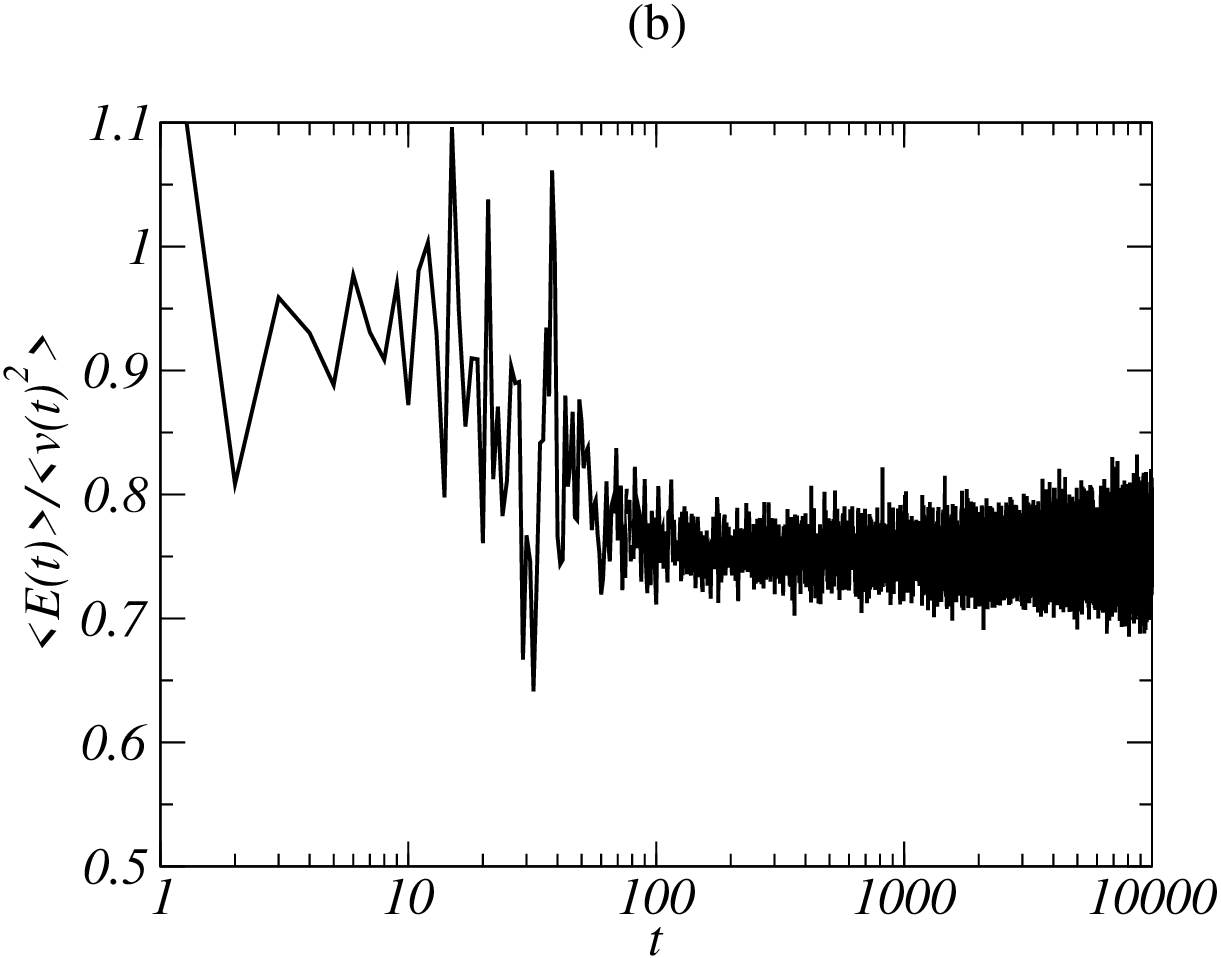}
\vspace*{1.0cm}
\includegraphics*[width=0.4\textwidth]{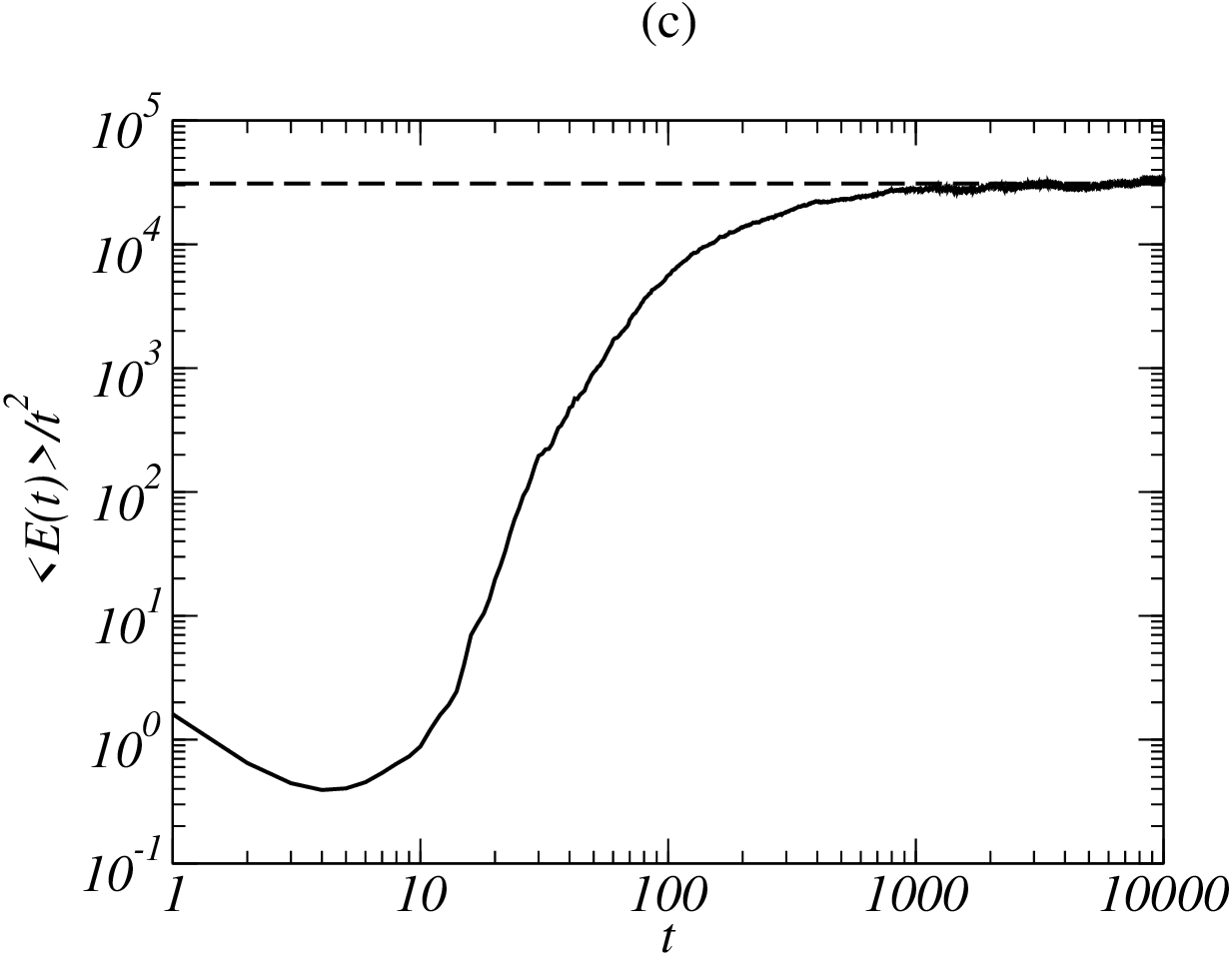}
\hfill
\includegraphics*[width=0.4\textwidth]{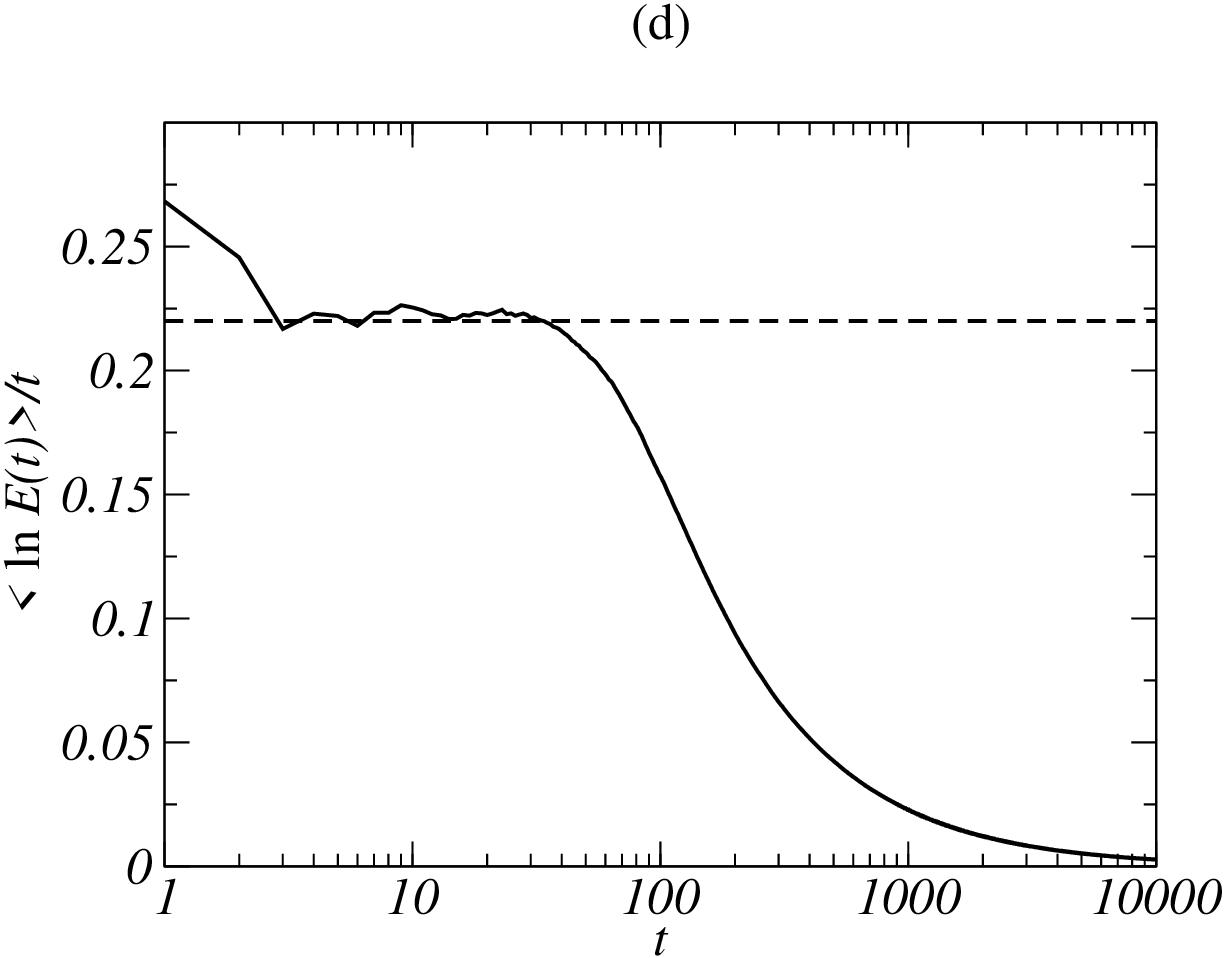}

    \caption{\label{fig:Duffing} Duffing oscillator: 
Eq.~(\ref{multnonlin}) is integrated numerically for $\omega = 1$,
$\lambda = 10^{-6}$ and ${\mathcal D} = 1$ with a timestep 
$\delta t = 10^{-4}$. Ensemble averages are
computed over $2 \, 10^3$ realizations.
Fig.~(a): $\langle E(t) \rangle$ vs. $t$; the dashed line corresponds
to the expected power law 
$\langle E(t) \rangle \simeq D_E(\lambda) \; t^2$.
Fig.~(b): the numerical value of the equipartition ratio 
$\langle E(t) \rangle/ \langle v(t)^2 \rangle$ changes from 
a value close to $1$ to $3/4$.
Fig.~(c): the numerical estimate of the generalized diffusion
constant $D_E(\lambda) = \lim_{t\rightarrow\infty} 
\langle E(t) \rangle/t^2$ is $3.0(4) \, 10^4$; the dashed line
corresponds to the expected diffusion constant 
$D_E(\lambda) = 3.13 \, 10^4$.
Fig.~(d): $\langle \ln E(t) \rangle/t$ vs. time $t$; the dashed line
corresponds to the Lyapounov exponent $\Lambda = 0.22$
expected in the linear regime.
}
\end{figure}

 When $\omega$ is non zero,  Eqs.~(\ref{solx})
 and (\ref{solv}) become:
 \begin{eqnarray}
          x &=& \left(  \frac{ 4E^2}{ 4E + \omega^4 }\right) ^{1/4}  
 \frac{ {\rm sn}\left((4E + \omega^4)^{\frac{1}{4}}t;k  \right)}
      { {\rm dn}\left((4E+ \omega^4)^{\frac{1}{4}}t;k \right)}  ,
  \label{dufx}  \\                 
     \dot x &=&  (2E)^{1/2}
 \frac{ {\rm cn}\left((4E + \omega^4)^{\frac{1}{4}}t;k \right)}
      { {\rm dn}^2\left((4E + \omega^4)^{\frac{1}{4}}t; k  \right)} ,
\label{dufv}
 \end{eqnarray}
where the elliptic modulus $ k$
  varies with  the energy and   is given by
  \begin{equation}
 k^2 = \frac{  \sqrt{4E + \omega^4}  - \omega^2 }
             { 2 \sqrt{4E + \omega^4}} .
 \label{modulus}
\end{equation}  
 We notice that $k$  tends to 
  the limiting value $1 / \sqrt{2}$ when the energy
  goes   to infinity. Defining  the  angle  variable  as:
\begin{equation}
  \phi =  {\rm sd}^{-1}\left(
\frac{(4E + \omega^4)^{\frac{1}{4}}}{ \sqrt{2 E}} \, x; \, k  \right) , 
\label{cubicangle2}
   \end{equation}
 we  rewrite the  dynamical  equation
  in energy-angle coordinates.
However, while deriving the dynamical equation for  $\phi$
  we must remember that the  elliptic modulus $k$ 
 depends on   the energy $E$ and is, therefore, a function of time.
  After  reintroducing   the  multiplicative  noise term,
  the stochastic Duffing
 oscillator    in the  energy-angle variables becomes:
  \begin{eqnarray}
 \dot E &=&   \frac { 2E } { (4E + \omega^4)^{\frac{1}{4}}} \,
  \frac{ {\rm sn}(\phi;k) {\rm cn}(\phi;k)   }
      { {\rm dn}^3(\phi;k)} \,  \xi(t)  \,\,  ;
\label{duffE}    \\
\dot\phi &=&  (4E + \omega^4)^{\frac{1}{4}} - \xi(t) 
 \frac{(2E + \omega^4) {\rm sd}^2(\phi;k)}{ (4E + \omega^4)^{\frac{5}{4}}}
   -  \omega^2 \xi(t) \left(  
\frac{  \, E \, {\rm sd}^4(\phi;k) }{ (4E + \omega^4)^{\frac{7}{4}}} \,
 -\frac{ E \,  {\rm sn}(\phi;k) {\rm cn}(\phi;k) 
   \int_0^\phi  {\rm sd}^2(\theta;k) d\theta   }
  {(4E + \omega^4)^{\frac{7}{4} }{\rm dn}^3(\phi;k) } \, \right)  \, .
 \label{duffangle}
\end{eqnarray}
 As compared to Eq.~(\ref{evolangle}), 
 two  supplementary terms  appear in Eq.~(\ref{duffangle}).
 These terms are  related to ${d k}/{d t}$ and  are
 proportional to $\omega^2$. 

Although the dynamical  equations
 are more complicated  than those of 
  the purely cubic case, the analysis can be performed
  as above. We shall  however
  simplify our discussion  here  
 by  taking $k$ equal to its 
 asymptotic value $1 / \sqrt{2}$.
 This approximation is  justified as soon as the energy is large.
 We also replace the noise $\xi(t)$   by the effective  noise $\Xi(t) $ 
 defined in   Eq.~(\ref{bruiteffect}). This second
 approximation is only qualitatively correct
 since it amounts to neglecting a  deterministic force 
 in the effective Langevin dynamics for $E$. We thus obtain:
\begin{equation}
 \dot E \simeq  \frac { 2E } { (4E + \omega^4)^{\frac{1}{4}}} \, \Xi(t) \, . 
\label{effecduff}
\end{equation}
 We deduce from    Eq.~(\ref{effecduff}) 
   that as long as  $E  \ll\omega^4$,
 the energy behaves as the exponential of a Brownian motion and, therefore,
 increases exponentially with time.  However,
 when  $ E > \omega^4 $,  the nonlinear term becomes important,
 Eq.~(\ref{duffE})  reduces to Eq.~(\ref{evolE}),
 and   the energy  grows  as  the square of time. 

 We expect that the   crossover from   exponential to  algebraic growth 
 will appear when $ E \sim \omega^4 $, or $x \sim \omega$. 
 Using unscaled variables, the balance between linear and non-linear terms 
 in  Eq.~(\ref{multnonlin}) is obtained when 
 $x = x_c \sim \omega/\sqrt{\lambda}$.  Fig.~\ref{fig:Duffing} 
 demonstrates  that the two regimes are observed numerically
 when the nonlinear coefficient $\lambda$ is very small compared
 to $\omega^2$: we use the numerical values $\omega = 1$ and 
 $\lambda = 10^{-6}$. We notice that in the short-time
 linear regime, the usual equipartition relation for a quadratic 
 potential is verified ($\langle E \rangle \simeq 
 \langle \dot x^2 \rangle$), while the exponential growth of the
  energy is characterized by the Lyapunov exponent $\Lambda$ defined in 
 Eq.~(\ref{deflambda}). In the long-time regime, the
 equipartition ratio reaches its nonlinear value $3/4$,
 while the energy growth becomes algebraic, with a generalized
 diffusion constant $D_E(\lambda)$ in good agreement with 
 Eq.~(\ref{Diffenerg}), up to the expected scaling factor: 
 $D_E(\lambda) = D_E/\lambda$.

  \section{The general nonlinear oscillator with parametric noise}
\label{sec:gen}

   We now consider the case of a particle trapped in a 
 confining potential ${\mathcal U}(x)$  and subject to an external
 noise. This mechanical system generalizes the Duffing
 oscillator studied in the previous section and its dynamics
 is given by
  \begin{equation}
   \frac{\textrm{d}^2 }{\textrm{d} t^2}x(t) + \xi(t)\, x(t) 
   = - \frac { \partial{\mathcal U}(x)}{\partial x} \,,
 \label{genU}
\end{equation}
 where  $\xi(t)$ is the Gaussian white noise of Eq.~(\ref{defgamma}).
 For the  potential to be confining, we must have 
  ${\mathcal U} \rightarrow +\infty$ when $ |x| \rightarrow \infty$. 
 We restrict our analysis to the case where ${\mathcal U}$
 is a polynomial;  in order to respect  the $ x \to -x$
 symmetry  ${\mathcal U}$ is     even in $x$.
 Hence, when $ |x| \rightarrow \infty$,
  \begin{equation}
     {\mathcal U}   \sim \frac{ x^{2n}}{2n}
  \,\, \hbox{ with } \,\, n \ge  2 \,, 
 \label{infU}
\end{equation}
 the coefficient of $x^{2n}$ is chosen to be 
  $1/2n$  by  a  suitable rescaling.
 As before, we expect the amplitude of the oscillator to grow without
 bounds at large times. Keeping only the relevant terms
 in   Eq.~(\ref{genU})  leads to
  \begin{equation}
   \frac{\textrm{d}^2 }{\textrm{d} t^2}x(t) + \xi(t)\, x(t) 
   + x(t)^{2n-1}  = 0 \,.
 \label{nthorder}
\end{equation}  

 The deterministic version of Eq.~(\ref{nthorder}),
   $\ddot x + x^{2n-1} = 0$,  is integrable because of
 energy conservation.
  For a given value of $E$ the deterministic solution is 
 \begin{eqnarray}
          x &=&   E^{1/{2n}} \, {\mathcal S}_n
 \left( (2nE)^{\frac{n-1}{2n}} t  \right) ,
  \label{solnx}  \\                 
     \dot x &=& (2n)^{\frac{n-1}{2n}} E^{1/2} \,
  {\mathcal S}_n'\left( (2nE)^{\frac{n-1}{2n}} t  \right) \, ,
\label{solnv}
 \end{eqnarray}
where the energy reads
 \begin{equation}
   E =  \frac{1}{2}\dot x^2 + \frac{1}{2n} x^{2n} \,.
\label{energyn}
\end{equation}
 The function ${\mathcal S}_n$
 is defined as the inverse function  of an 
  hyperelliptic integral as follows
 \begin{equation}
{\mathcal S}_n(X) = Y \,  \leftrightarrow 
  X =  \sqrt{n} \int_0^{ \frac{Y}{(2n)^{1/2n}}} 
 \frac{{\textrm d}u}{\sqrt{ 1 - u^{2n}}}
  =  \frac{ \sqrt{n}} { (2n)^{1/2n} } \int_0^Y 
   \frac{{\textrm d}u}{\sqrt{ 1 -  \frac{u^{2n}}{2n}}}      \,    .
  \label{hyperelli}
\end{equation}
 From this definition, we find a relation between  ${\mathcal S}_n$ 
and its derivative ${\mathcal S}_n'$
\begin{equation}
 {\mathcal S}_n'(X) = \frac{ (2n)^{ \frac{1}{2n}}}{\sqrt{n}} \left(
      1 -  \frac{({\mathcal S}_n(X))^{2n}}{2n} \right)^{\frac{1}{2}} .
\label{derivS}
\end{equation}
 The  action-angle variables for the deterministic
 system  are  given by
\begin{eqnarray}
        I &=& 4 \int_0^{(2nE)^{\frac{1}{2n}}} 
\sqrt{ 2E - \frac{x^{2n}}{n}} {\textrm d}x
  = 4 \, (2^{n+1}n)^{\frac{1}{2n}} \, E^{ \frac{n+1}{2n} }
 \int_0^1  \sqrt{ 1 - u^{2n}} {\textrm d}u  
 \propto E^{ \frac{n+1}{2n} } \,   ,  \\
      \phi &=& {\mathcal S}_n^{-1} \left(\frac{x}{ E^{\frac{1}{2n}} }\right)
  =  \sqrt{n} \, \int_0 ^{ \frac{x}{(2nE)^{{1}/{2n}}} }
    \frac{{\textrm d}u}{\sqrt{ 1 - u^{2n}}} \, .
\end{eqnarray}
 The angle  variable  $\phi$  is well
 defined modulo the period  $4 K_n$ of the function
 ${\mathcal S}_n$  where
\begin{equation}
   K_n   = \sqrt{n} \int_0^1  \frac{{\textrm d}u}{\sqrt{ 1 - u^{2n}}} \, .
\label{nperiod}
\end{equation}

\begin{figure}[ht]
\hfill
\includegraphics*[width=0.4\textwidth]{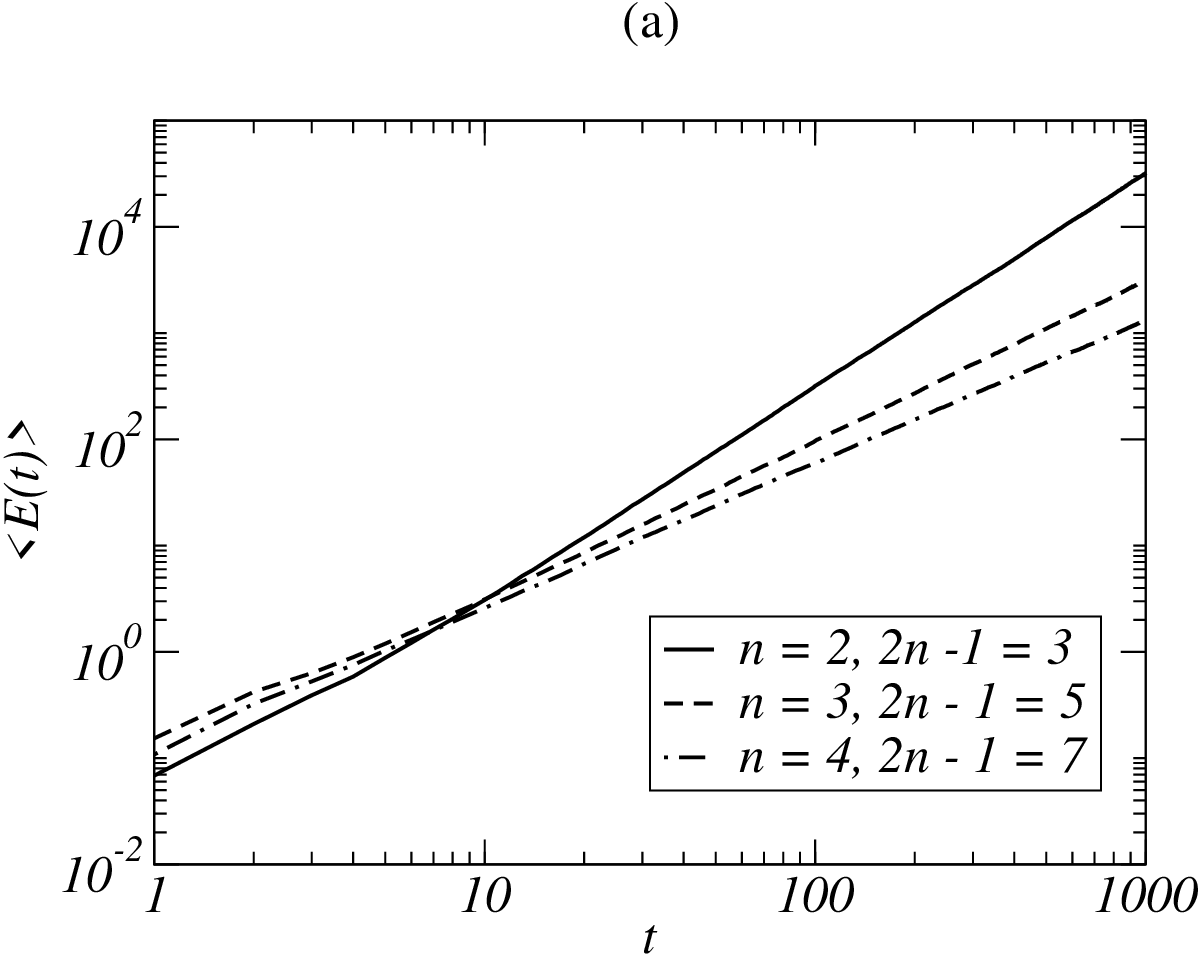}    
\hfill
\includegraphics*[width=0.4\textwidth]{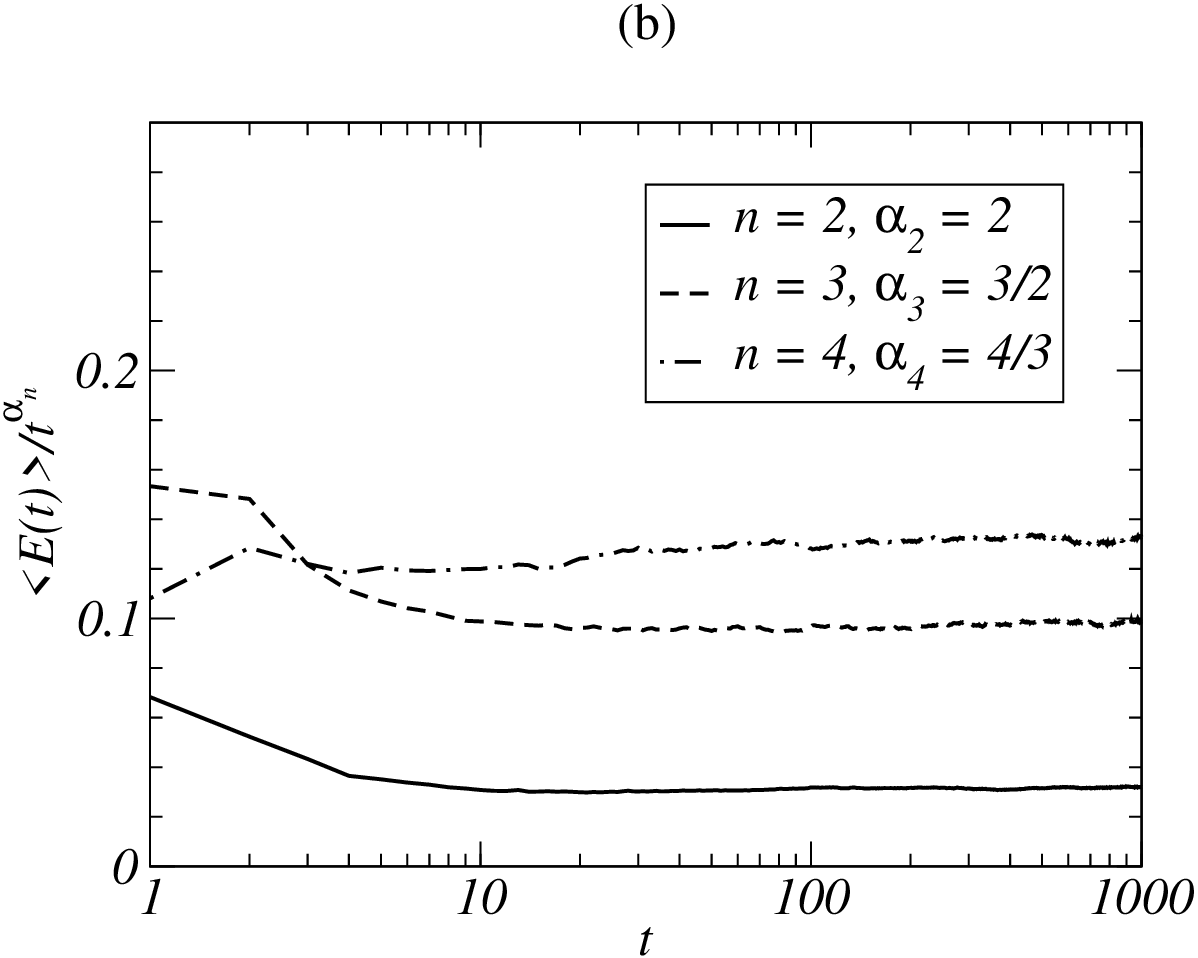}
\hfill
\vspace*{0.1cm}
\hfill
\includegraphics*[width=0.4\textwidth]{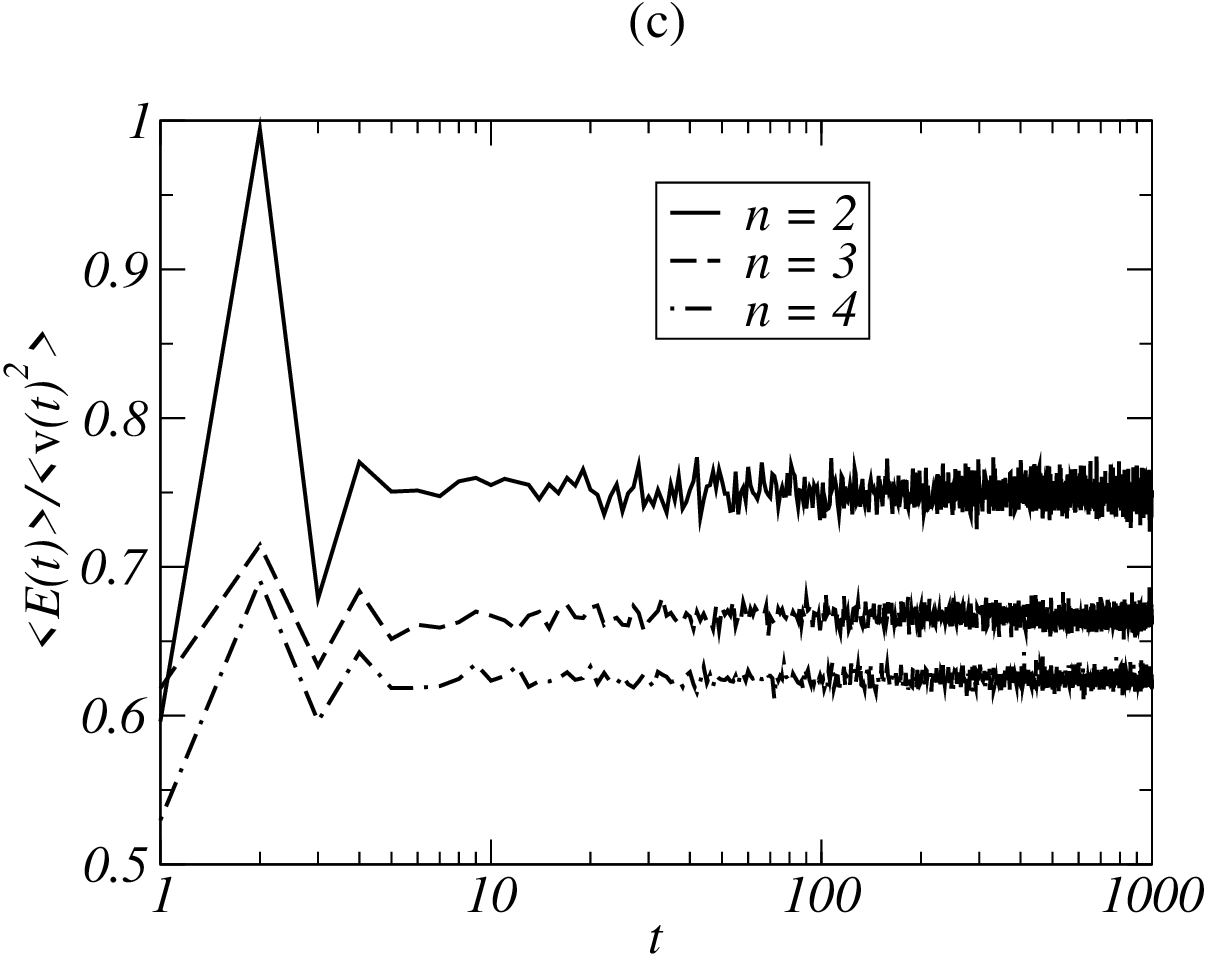}
\hfill

    \caption{\label{fig:genU} General nonlinear oscillator: 
Eq.~(\ref{nthorder}) is integrated numerically for ${\mathcal D} = 1$, 
with a timestep $\delta t$, and averaged over $10^4$ realizations for
$n=2$ ($2n - 1 = 3$), $\delta t = 5 \, 10^{-4}$;
$n=3$ ($2n - 1 = 5$), $\delta t = 5 \, 10^{-4}$;
$n=4$ ($2n - 1 = 7$), $\delta t = 10^{-4}$.
Fig.~(a): average energy $\langle E(t) \rangle$ vs. time $t$.
Fig.~(b): the limit $\lim_{t \rightarrow \infty} 
\langle E(t) \rangle/t^{\alpha_n}$, $\alpha_n = \frac{n}{n-1}$,
yields the following estimates of the diffusion constants
$D_E^{(n)}$: $D_E^{(2)} = 0.031(1)$, 
$D_E^{(3)} = 0.097(3)$, $D_E^{(4)} = 0.130(5)$,
in excellent agreement with the predictions of
 Eqs.~(\ref{penta}) and (\ref{hepta}).
Fig.~(c): the measured equipartition ratio 
$\langle E(t) \rangle / \langle v(t)^2 \rangle$ is close to
the theoretical value $\frac{n+1}{2n}$  given in  Eq.~(\ref{equipn2}):
$3/4$ for $n=2$; $2/3$ for $n=3$; $5/8$ for $n=4$.
Data for $n=2$ is the same as that of  Figs.~\ref{fig:cubic:equi} and
\ref{fig:cubic:scaling}.
}
\end{figure}

 When the noise term is taken into account, the energy is not conserved
  and the stochastic evolution of the variables $E$  and $\phi$
 is given by
\begin{eqnarray}
\dot{E}   &=&   x \dot{x} \xi(t) = 
 (2n)^{\frac{n-1}{2n}} \, E^{\frac{n+1}{2n} } \,
  {\mathcal S}_n(\phi) {\mathcal S}_n'(\phi)  \, \xi(t) 
\label{evolnE}   \, ,  \\ 
   \dot\phi &=& (2nE)^{\frac{n-1}{2n}} - \frac{1}{ (2n)^{\frac{1}{n}}} \,
   \frac{  {\mathcal S}_n(\phi)^2} { (2nE)^{\frac{n-1}{2n}} } \, \xi(t)
\label{evolnphi}  \, .
   \end{eqnarray}
  Introducing the  auxiliary variable $\Omega$,  
 \begin{equation}
 \Omega =   (2n)^{ \frac{n+1}{2n} } \,  E^{\frac{n-1}{2n}}  \, ,
  \end{equation}
  the  Eqs.~(\ref{evolnE}) and (\ref{evolnphi}) can be written 
   in  the simpler form:
   \begin{eqnarray}
     \dot \Omega  &=& (n -1) \, {\mathcal S}_n(\phi)
 {\mathcal S}_n'(\phi) \, \xi(t)   \label{evoln1}
   \,  ,   \\    \dot\phi  &=& \frac{\Omega}{ (2n)^{\frac{1}{n}}}
 - \frac{{\mathcal S}_n(\phi)^2}{\Omega}  \, \xi(t)  \, .
    \label{evoln}
   \end{eqnarray}
 As in the case of the Duffing oscillator, we observe that  $\phi$
   is a fast variable and therefore we shall
   average out  its rapid variations.
   Starting from (\ref{solnv}), we  write
\begin{equation}
  \langle \dot x^2   \rangle = (2n)^{\frac{n-1}{n}} \,
   \overline { {\mathcal S}_n'(\phi)^2}  \,  \langle  E   \rangle  
 =   2\, \frac{  \int_0^1 \textrm{d}u \, \sqrt{ 1 - u^{2n}}  } 
     {  \int_0^1 \frac{\textrm{d}u}{\sqrt{ 1 - u^{2n}} } } 
  \, \langle  E   \rangle \,  ,
\label{equipn1}
\end{equation}
the last equality is derived by writing  $u = {\mathcal S}_n(\phi)$,
 and  using  Eqs.~(\ref{derivS}) and (\ref{nperiod}). Moreover,  the
  following identity is true  (as can be shown  by integrating
 $ \int_0^1  1. \sqrt{ 1 - u^{2n}} \, \textrm{d}u$ by parts): 
 \begin{equation}
   \int_0^1 \textrm{d}u \, \sqrt{ 1 - u^{2n}} = n \int_0^1
\textrm{d}u \, \frac{u^{2n} }{\sqrt{ 1 - u^{2n}} }  = 
 -n \int_0^1 \textrm{d}u \, \sqrt{ 1 - u^{2n}}
 +n  \int_0^1 \frac{\textrm{d}u}{\sqrt{ 1 - u^{2n}} } \, .
\label{ipp}
\end{equation} 
 Substituting this identity in Eq.~(\ref{equipn1}) leads to
\begin{equation}
  \langle  E   \rangle   = \frac{n+1}{2n} \, \langle \dot x^2   \rangle .
\label{equipn2}
\end{equation}
 From  the definition (\ref{energyn}) of the energy, we derive 
 another  statistical equality:
\begin{equation}
    \langle \dot x^2   \rangle  = \langle  x^{2n}   \rangle  \, .
\label{equipnxv}
\end{equation}
 We  emphasize again  that these generalized equipartition
  relations are valid for
 any type of noise,  colored or white, additive or 
 multiplicative. The only hypothesis is that the probability
  distribution function $P_t(\Omega,\phi)$ becomes 
  uniform in $\phi$ over the interval $[0, 4 K_n]$
 when $ t \rightarrow \infty$. We observe from the numerical simulations
 presented in Fig.~\ref{fig:genU}.c  that this condition is very well 
 satisfied.

 The   averaging  process  with respect to the fast variable 
 generates  an effective Langevin
 dynamics for  the  slow  variable $\Omega(t)$. Starting
 from the Fokker-Planck equation for the total P.D.F., 
 $P_t(\Omega,\phi),$  and averaging over $\phi$ 
 leads  to the following evolution equation for the
 averaged probability distribution   ${\tilde P}_t(\Omega)$
\begin{equation}
   \partial_t {\tilde P}  =  \frac{{\mathcal M}_n {\mathcal D} }{2} 
\left(  \partial_{\Omega}^2 {\tilde P} - \frac{2}{n-1} \, \partial_{\Omega}
      \frac{{\tilde P}}{\Omega}  \right)  \, ,
\label{nFPmoy}
\end{equation}
 where ${\mathcal M}_n$ is given by
\begin{equation}
    {\mathcal M}_n  = \frac { (n-1)^2}{n} \, (2n)^{\frac{2}{n}} \,
           \frac{\int_0^1 {\textrm d}u \, u^2\sqrt{ 1 - u^{2n} } }
                       { \int_0^1 \frac{{\textrm d}u} {\sqrt{ 1 - u^{2n} }}}
  =   \frac { (n-1)^2}{n+1}   \, (2n)^{\frac{2}{n} } \,
    \frac {\Gamma(\frac{3}{2n})}{\Gamma(\frac{1}{2n})} \,
    \frac{\Gamma(\frac{3n+1}{2n})}{\Gamma(\frac{3n+3}{2n})}  \, ,
\label{nmeanval}
\end{equation}
  $\Gamma(.)$   being  the Euler Gamma function. 
  The  effective
 Langevin dynamics for the variable $\Omega$  is thus 
\begin{equation}
   \dot\Omega =  \frac{{\mathcal M}_n \, {\mathcal D} }{n-1}
  \, \frac{1}{\Omega}  + \Xi_n(t)  \, ,
\label{nLangeff} 
\end{equation}  
where the effective Gaussian white
 noise $\Xi_n(t)$  satisfies the relation 
 \begin{equation}
\langle  \Xi_n(t) \, \Xi_n(t') \rangle = 
  {\mathcal M}_n  {\mathcal D} \, \delta( t  - t')  \, .
  \label{nbruiteffect}
 \end{equation}

  The  averaged  distribution
 fonction ${\tilde P}_t(\Omega)$ can  be  calculated  because
 Eq.~(\ref{nFPmoy}) is exactly solvable due to its  invariance 
 under rescalings $ t \rightarrow \lambda^2  t $ , 
  $ \Omega  \rightarrow \lambda \Omega \, , $  $ \lambda $  being  an arbitrary
 real number. The    P.D.F. of  $ \Omega $ is found to be 
 \begin{equation}
   {\tilde P}_t(\Omega) =
  \frac{1}{ \Gamma \left(\frac{n + 1}{2 (n-1)}\right)} \,
\frac{\Omega^{\frac{2}{n-1} }}
     {\left(2 {\mathcal M}_n {\mathcal D}   t \right)^
              {\frac{n + 1}{2 (n-1)}}}  \,
 \exp\left\{ - \frac{\Omega^2}{2 {\mathcal M}_n {\mathcal D} t } \right\} \, ,
\label{pdf2Om}
\end{equation}
 from which we obtain the P.D.F. of the energy:
 \begin{equation}
   {\tilde P}_t(E) =
  \frac{1}{  \Gamma \left(\frac{n + 1}{2 (n-1)}\right)} \,
\frac{n-1}{n E} \,
\left( \frac{(2n)^{\frac{n+1}{n}} \; E^{\frac{n-1}{n}}}
{2 {\cal M}_n {\cal D} t}\right)^{\frac{n+1}{2 (n-1)}}
\exp \left\{ - 
\frac{(2n)^{\frac{n+1}{n}} \; E^{\frac{n-1}{n}}}
{2 {\cal M}_n {\cal D} t}
 \right\} \, .
\label{pdf2E}
\end{equation}

 The long-time behavior of the amplitude,
 velocity and energy of the general nonlinear oscillator can  
 now be derived.
 Using the equipartition identity (\ref{equipn2})
  and Eq.~(\ref{pdf2E}), we obtain
 \begin{eqnarray}
 \langle E  \rangle 
   &=& \frac{1}{(2n)^{\frac{n+1}{n-1}}} \frac{\Gamma\left( \frac{3n+1}{2n-2}
   \right)}  { \Gamma\left( \frac{n+1}{2n-2} \right)}  
\left( 2 {\mathcal D} {\mathcal M}_n  t \right)^{\frac{n}{n-1}} 
\propto t^{\frac{n}{n-1}} \,  ,
 \label{moynE} \\
     \langle \dot{x}^2  \rangle  &=& \frac{2n}{n+1}  \langle E  \rangle 
\propto t^{\frac{n}{n-1}}  \, .
\label{moynv}
  \end{eqnarray}
Using Eq~(\ref{nmeanval}), we find that 
\begin{equation}
   \overline{ {\mathcal S}_n^2(\phi) } =  \frac{1}{K_n}
   \int_0^{K_n} {\mathcal S}_n^2(\phi)\, {\textrm d}\phi = 
  (2n)^{\frac{1}{n}} 
\frac{ \int_0^1 \frac{u^2{\textrm d}u} {\sqrt{ 1 - u^{2n} } } }
    { \int_0^1 \frac{{\textrm d}u} {\sqrt{ 1 - u^{2n} } }  }
  =  \frac{ n + 3} { (2n)^{\frac{1}{n}} (n-1)^2 }  {\mathcal M}_n  \, ,
\label{vals2phi}
\end{equation}
where we made the change of variables   $ u = {\mathcal S}_n(\phi) ,$
 and used  the following identity (obtained by integrating by parts):
 \begin{equation}
 \int_0^1 u^2\sqrt{ 1 - u^{2n} } \, {\textrm d}u
           = \frac{n}{n+3} \,
 \int_0^1 \frac{u^2{\textrm d}u} {\sqrt{ 1 - u^{2n} }} \, . 
\end{equation}
  Finally,   we deduce  from  Eqs.~(\ref{solnx}), (\ref{pdf2E})
 and (\ref{vals2phi})
\begin{equation}
 \langle x^2  \rangle  =
 \overline {{\mathcal S}_n^2(\phi) }  \, \langle E^{1/n} \rangle
 =  \frac{2} {n-1} \, \frac{{\cal M}_n}{(2n)^{\frac{1}{n}}} \,
\frac{\Gamma\left( \frac{3n+1}{2n-2}   \right)} 
 { \Gamma\left( \frac{n+1}{2n-2} \right)} \, 
\left( \frac{2 {\mathcal M}_n {\mathcal D}  t}{(2n)^{\frac{n+1}{n}}} \right)^
 {\frac{1}{n-1}} \propto t^{\frac{1}{n-1}}   \,  .
 \label{moynx}
\end{equation}

  The analytical results   for the nonlinear oscillators  with 
  quintic $x^5$ ($n = 3$) and  heptic nonlinearity  $x^7$ 
  ($n = 4$) are as follows:
 \begin{eqnarray}
   &\hbox{For $ n =3 $},& \,\,  \langle E  \rangle  = 0.097 \,t^{3/2} \,\,\, , 
  \,\,\,  \langle \dot{x}^2  \rangle  =   0.145 \,t^{3/2}  \,\,\,  ,
   \,\,\,  \langle x^2  \rangle     = 0.290 \, t^{1/2}     \,\, .
 \label{penta}   \\
  &\hbox{For $ n =4 $},& \,\, \langle E  \rangle  = 0.130 \,t^{4/3} \,\,\, ,
  \,\,\, \langle \dot{x}^2  \rangle  = 0.208  \,t^{4/3}  \,\,\,   ,  
   \,\,\,  \langle x^2  \rangle  =   0.347 \,t^{1/3}      \,\, . 
\label{hepta}
\end{eqnarray}
The formulas (\ref{moynE}), (\ref{moynv}) and
 (\ref{moynx}) were verified numerically. The scaling exponents
 and the prefactors given in Eqs.~(\ref{penta}) and (\ref{hepta})
 are in excellent agreement with the numerical values, as shown
in Fig.~\ref{fig:genU}.

In conclusion,  we have derived  the  following scaling relations:
\begin{eqnarray}
          E      &\sim&  t^{\frac{n}{n-1}} ,  \nonumber   \\ 
          x      &\sim&  t^{\frac{1}{2(n-1)}} , \nonumber    \\ 
         \dot x  &\sim&  t^{ \frac{n}{2(n-1)} }  .
\label{scalings2}
\end{eqnarray}
  In particular, it should be remarked  that $x$ 
 undergoes an anomalous  diffusion   with time with
 exponent $1/{(2n-2)}$. If we make $n \to 1$ formally,
  this exponent diverges to infinity:
 this is consistent with the exponential growth
 of the linear oscillator (see Sec.~\ref{sec:linear}).

\hfill\break

 We end this section by considering the case of a general
  confining  potential energy  ${\mathcal U}$ neither  necessarily  polynomial
 in $x$ nor even in  $x$. The only requirement is that 
${\mathcal U} \rightarrow +\infty$ when $|x| \rightarrow \infty $.
 We discuss  the  qualitative  behavior of $E$, $\dot x$ and $x$ 
  at large times  from  elementary scaling  considerations.
 Suppose  first that ${\mathcal U} \sim |x|^{r}$  for
 large values of $|x|$,  $r$  being  an arbitrary real number. 
\begin{itemize}
 \item If $ r > 2$, then balance  between kinetic and potential energies 
 leads to  $E \sim \dot x^2  \sim x^r $; thus   the
 time evolution of the energy  is given by  $\dot E \sim  x \dot x \xi
  \sim E^{\frac{1}{r} -\frac{1}{2}} \xi$.  From 
 the scaling relations   $\dot E \sim E/t$
 and  $\xi \sim t^{-1/2} $, we conclude that
$$  E    \sim  t^{\frac{r}{r-2}} ,  \,\,\,\,  
    x    \sim   t^{\frac{1}{r-2}} ,  \,\,\,\,  
         \dot x  \sim  t^{\frac{r}{2(r-2)} } \,\, . $$
This qualitative argument can be made rigorous by generalizing
the results obtained above to non-integer values of $r$.
   \item If $ r \le  2$, then the potential  ${\mathcal U}$
 is negligible with respect to the multiplicative noise term 
 and we are back to the case of the degenerate linear oscillator. Therefore
 $E$, $\dot x$ and $x$ grow exponentially with time.
 \end{itemize}
 If  the potential 
 grows exponentially, {\it i.e.}  ${\mathcal U} \sim {\rm e}^{x^{\beta}}$,
  $\beta$  being   a positive real number, then  similar considerations
 lead to $ E \sim t $ and $\dot x \sim t^{1/2}$ 
 (leaving aside  logarithmic corrections). We then conjecture that the
  amplitude $x$  diffuses  in a logarithmically 
 slow manner:  $x  \sim (\ln t )^{1/\beta} $.

  \section{Colored  Gaussian  noise}
\label{sec:color}

  We now  consider the case where the Gaussian noise has a  
non-zero  correlation time, and  discuss
 how the previously   found  scalings  are modified.
 The system we want to study satisfies  the dynamical equation:
\begin{equation}
   \frac{\textrm{d}^2 }{\textrm{d} t^2}x(t) + x(t) \, \eta(t)
+ x(t)^{2n-1} = 0 ,
 \label{colorn}
\end{equation}
  where $\eta$ is a colored Gaussian noise  of zero mean-value.
 The statistical properties of  $\eta$  are determined by
 \begin{eqnarray}
       \langle \eta(t)  \rangle &=&   0 \, ,\nonumber \\
   \langle \eta(t) \eta(t') \rangle  &=&   
\frac{\mathcal D}{2 \, \tau}   \, {\rm e}^{-|t - t'|/\tau} \,,
   \label{deftau}
 \end{eqnarray} 
where $\tau$ is the correlation time of the noise. The noise $\eta$
 can be obtained from  white noise  by  solving the Ornstein-Uhlenbeck
 equation:
  \begin{equation} 
 \frac{{\textrm d} \eta(t)}{{\textrm d} t} = -\frac{1}{\tau} \eta(t) + 
\frac{1}{\tau} \xi(t),
  \label{OU}
\end{equation} 
 where $\xi(t)$ is the Gaussian white noise defined in Eq.~(\ref{defgamma}),
and $t, t' \gg \tau$.

\begin{figure}[th]
\hfill
\includegraphics*[width=0.4\textwidth]{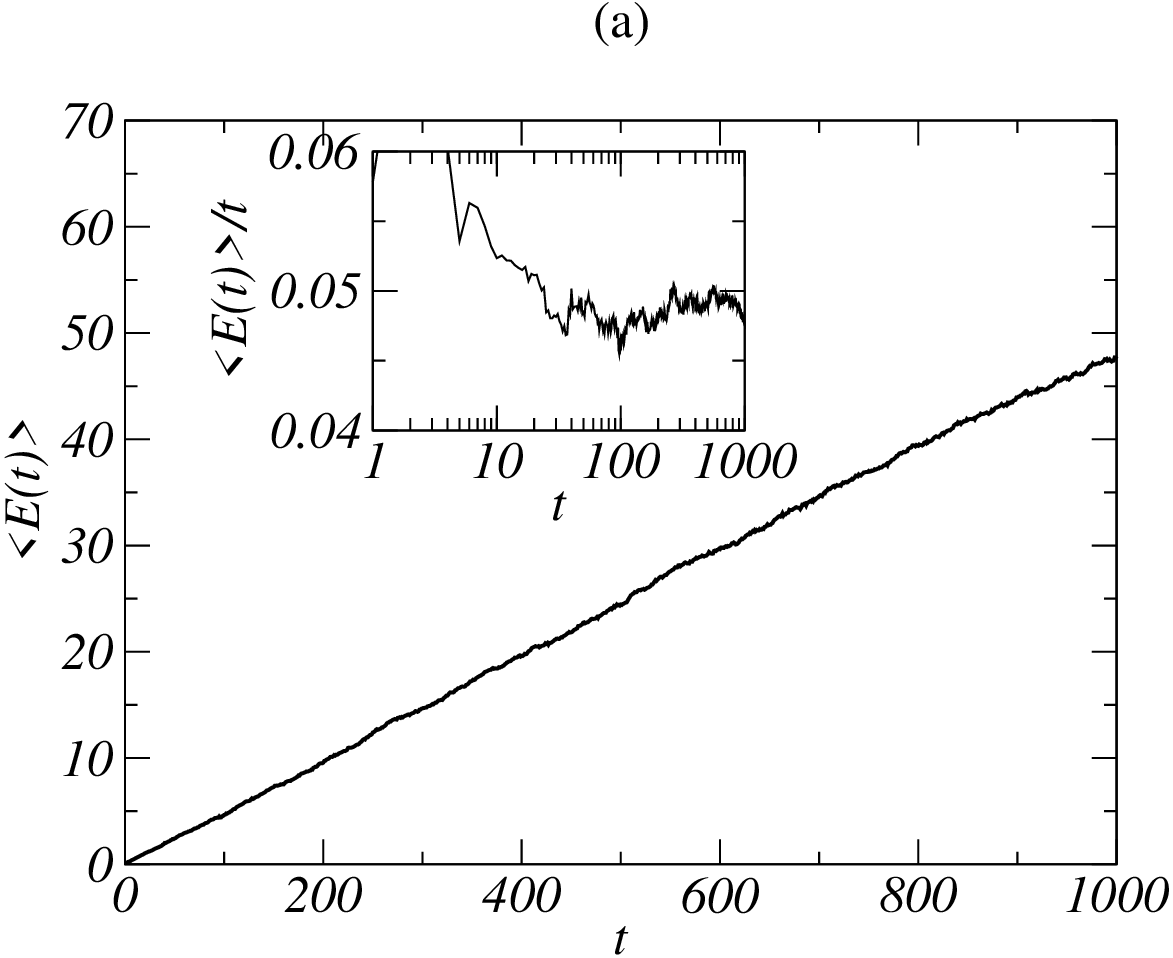}    
\hfill
\includegraphics*[width=0.4\textwidth]{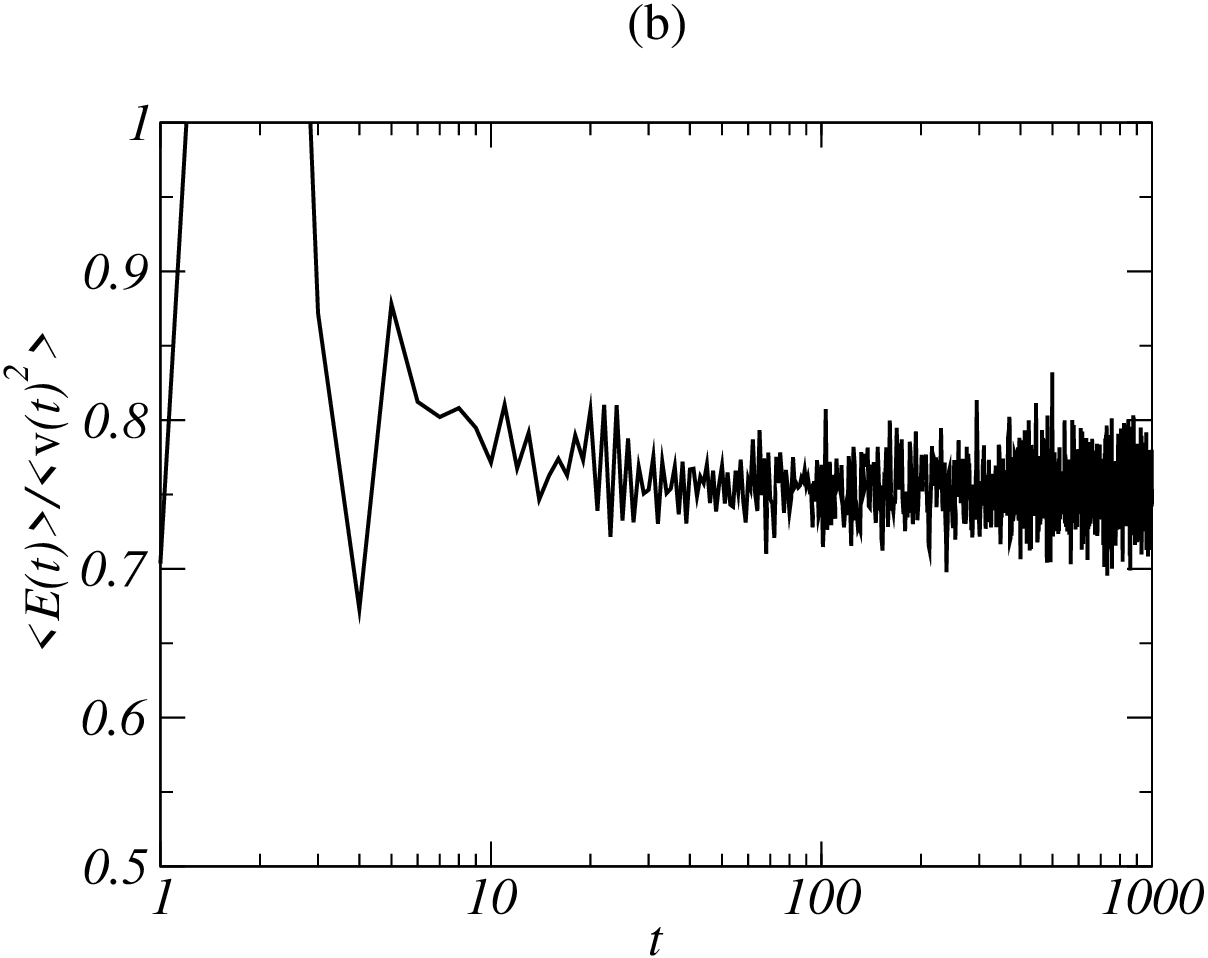}    
\hfill

    \caption{\label{fig:color} Cubic oscillator with
colored noise.
Eqs.~(\ref{colorn},\ref{OU}) are integrated numerically for $n = 2$,
$2 n -1  = 3$, ${\mathcal D} = 1$,
$\tau = 1$, with a timestep $\delta t = 10^{-5}$. Ensemble averages are
computed over $10^3$ realizations.
Fig.~(a): the average energy $\langle E(t) \rangle$ grows 
linearly with time. The inset gives an estimate of the
diffusion constant $D_E \simeq 0.047(3)$. 
Fig.~(b): the time-asymptotic value
of the equipartition ratio $\langle E(t) \rangle / \langle v(t)^2 \rangle$
is equal to $3/4$.
}
\end{figure}

Introducing action-angle variables as in Sec.~\ref{sec:gen}, we
obtain the same set of coupled Langevin equations, 
Eqs.~(\ref{evoln1},\ref{evoln}),
where $\xi(t)$ is replaced by the colored noise $\eta(t)$.
As emphasized previously, the generalized equipartition relations
are independent of the nature of the noise: Eqs.~(\ref{equipn2}) and 
(\ref{equipnxv}) remain valid when the noise is correlated in time.
This is confirmed by numerical simulations (see Fig.~\ref{fig:color}.b
for $n = 2$).

  The scalings found in sections \ref{sec:duff} and \ref{sec:gen} are  
  deduced  by  averaging
 the Fokker-Planck equation. Here, we must  write
 the evolution equation for the joint P.D.F. of $x$, $v=\dot{x}$
 and  $\eta$, $P_t(x,v,\eta)$:
\begin{equation}
\frac{ \partial P}{\partial t} = - v\frac{\partial P}{\partial x}
 +  (x^{2n-1} - x\eta)\frac{ \partial P}{\partial v}
 + \frac{1}{\tau} \frac{ \partial \eta P}{\partial \eta}
 +\frac{{\mathcal D}}{2 \tau^2}\frac{ \partial^2 P}{\partial \eta^2}\,  .
\label{FPcolor}
\end{equation}
 We  perform  a scaling analysis of this equation in the
 spirit of \cite{yp}.  Balancing  the diffusion term
 with  the time derivative leads to $\eta \sim t^{1/2}$.
 Then we compare  the terms of probability current
  $\frac{vP}{x}$  and  $\frac{(x^{2n-1} - x\eta)P}{v} . $ 
 A consistent balance between these terms is  possible only  if 
 $v^2 \sim x^{2n}$ and $x^{2n-2} \sim \eta$. We thus  find
 the  following scaling relations 
\begin{eqnarray}
          E      &\sim&  t^{\frac{n}{2(n-1)}} ,  \nonumber   \\ 
          x      &\sim&  t^{\frac{1}{4(n-1)}} , \nonumber    \\ 
         \dot x  &\sim&  t^{\frac{n}{4(n-1)} }  .
\label{scalingcolor}
\end{eqnarray}
  Thus, we  predict  that the scaling exponents for colored
 noise are {\it half} the exponents calculated
 for white noise (\ref{scalings2}). 
Numerical simulations indeed confirm that the average energy
of a cubic oscillator ($n = 2$) with colored multiplicative 
noise grows linearly with time (see Fig.~\ref{fig:color}.a).

 The period $T$ of  a  deterministic oscillator
    (without noise)   decreases as the energy increases:  $T \sim 
 E^{-\frac{n-1}{2n}}$ from Eq.~(\ref{evolnphi}).  When  the equations
 are written in the energy-angle coordinates,
   two time scales
 $T$  and $\tau$  appear. In the regime where $\tau \ll T$,
 the correlation time of the noise is much smaller than
 the typical variation time of the angle. Hence  the noise 
 can be considered to be  white, the averaging procedure can be applied
 as in Section \ref{sec:gen} 
and the scalings found in  (\ref{scalings2}) are correct.
 When $T \sim \tau$ the noise becomes correlated
 over a period of the free system and can not be treated as white anymore.
  Now, $T \sim \tau$  corresponds to  $E \sim \tau^{-\frac{2n}{n-1}}$ which
 leads to  crossover time $t_c$ of the order
  $t_c^{\frac{n}{n-1}}  \sim \tau^{-\frac{2n}{n-1}}$  {\it i.e.}  
 $ t_c \sim  \tau^{-2}$. 
 For times larger than $t_c$, the scalings (\ref{scalingcolor})
 are observed.

   \section{Conclusion}
\label{sec:conc}

         A particle trapped in a confining potential with white  multiplicative
   noise undergoes anomalous diffusion: if the confining potential 
 grows as  $x^{2n}$ at infinity,  the particle diffuses as
 $D_xt^{\beta_n}$. We have calculated 
  the anomalous diffusion  exponent  $\beta_n = 1/(2n -2)$, and
  the  coefficient $D_x$.  We have found similar laws for the
  diffusion of velocity and energy. Thanks to generalized
  equipartition identities, we  have derived  universal relations
  between the exponents and between  the prefactors. Our calculations
 are based on the assumption that in the long-time limit
 the probability distribution function becomes uniform in the phase
 variable.  By averaging out the phase variations,
 an effective, projected, dynamics for the action (or energy) can
 be defined. This technique  enabled us to derive  the  asymptotic distribution
 law of the energy in  the $t \to \infty$ limit, 
 and to calculate  its non-Gaussian features (skewness and flatness).
 Our analytical results 
 agree  with  the  numerical  simulations  within  the numerical
   error bars. Thus, the averaging procedure  produces  very 
 accurate  results;  it  would be 
  an  interesting and  challenging problem  to  characterize  deviations
 from our results and  to  calculate  subleading  corrections. 

 In the case of colored  multiplicative  noise, we have deduced 
 the anomalous diffusion  exponents  from an elementary
 scaling argument. Our result, supported by  numerics,  shows 
 that the exponents are halved in presence of time correlations.
 The efficiency of parametric amplification decreases if
 the noise is coherent over  a period of the system 
 and therefore   the particle  diffuses at a much slower rate.
  In this case, however, the
  averaging technique is harder to apply  because 
 the  noise itself  is  averaged out  to  the leading order. 
 A precise calculation in the case of colored Gaussian  noise 
 still remains to be done.

 We have considered only  Hamiltonian systems, {\it i.e.} systems
 where no friction is present. Nevertheless, if the damping 
 is small,  the results we have derived  for the undamped  oscillator
  remain valid until the  crossover time (identical to 
  the typical decay time of the energy)   is reached. 
 The  general case   of a non-linear oscillator
 with (linear) friction   leads to
 interesting results and  is currently under study \cite{phkir}.

\acknowledgements

  It is a pleasure to thank Yves Pomeau for encouraging us to work
  in this  field of nonlinear stochastic equations and for his
  advice. K.M. is grateful to Michel Bauer for many useful discussions.

\appendix

  \section{Nonlinear oscillator with external and internal noise}
\label{sec:add}
   
 In this Appendix  we discuss 
 the behavior of an oscillator submitted to  both
  additive  and multiplicative  noises. 
  Because  we are considering non-dissipative systems, there is 
 no stationary probability distribution;  the position,
 velocity and energy of the system  satisfy scaling laws.

  We first consider the case where the noise is only additive.
 The dynamics of a  linear
 oscillator with additive noise is exactly solvable and 
 the solution of  the equation:
\begin{equation}
 \frac{\textrm{d}^2 }{\textrm{d} t^2}x(t) + \omega^2 x(t) =  \xi(t) ,
 \label{addlin}
\end{equation}
 is given by:
 \begin{equation}
  x(t) = \frac{1}{\omega} \, 
\int_0^t \sin \left(\omega(t - s)\right) \, \xi(s) \, {\textrm d}s \, .
\label{addsol}
\end{equation}
 We  conclude from this  expression  
 that $x(t)$ grows diffusively with time:
\begin{equation}
    \langle  x^2(t) \rangle =  \frac{{\mathcal D}}{2\omega^2}\left(
    t - \frac{\sin 2 \omega t}{ 2 \omega } \right) .
\end{equation}
 Such a diffusive behavior  is entirely  different from the exponential
 growth of a linear oscillator subject to parametric noise.

\begin{figure}[t]
\hfill
\includegraphics*[width=0.4\textwidth]{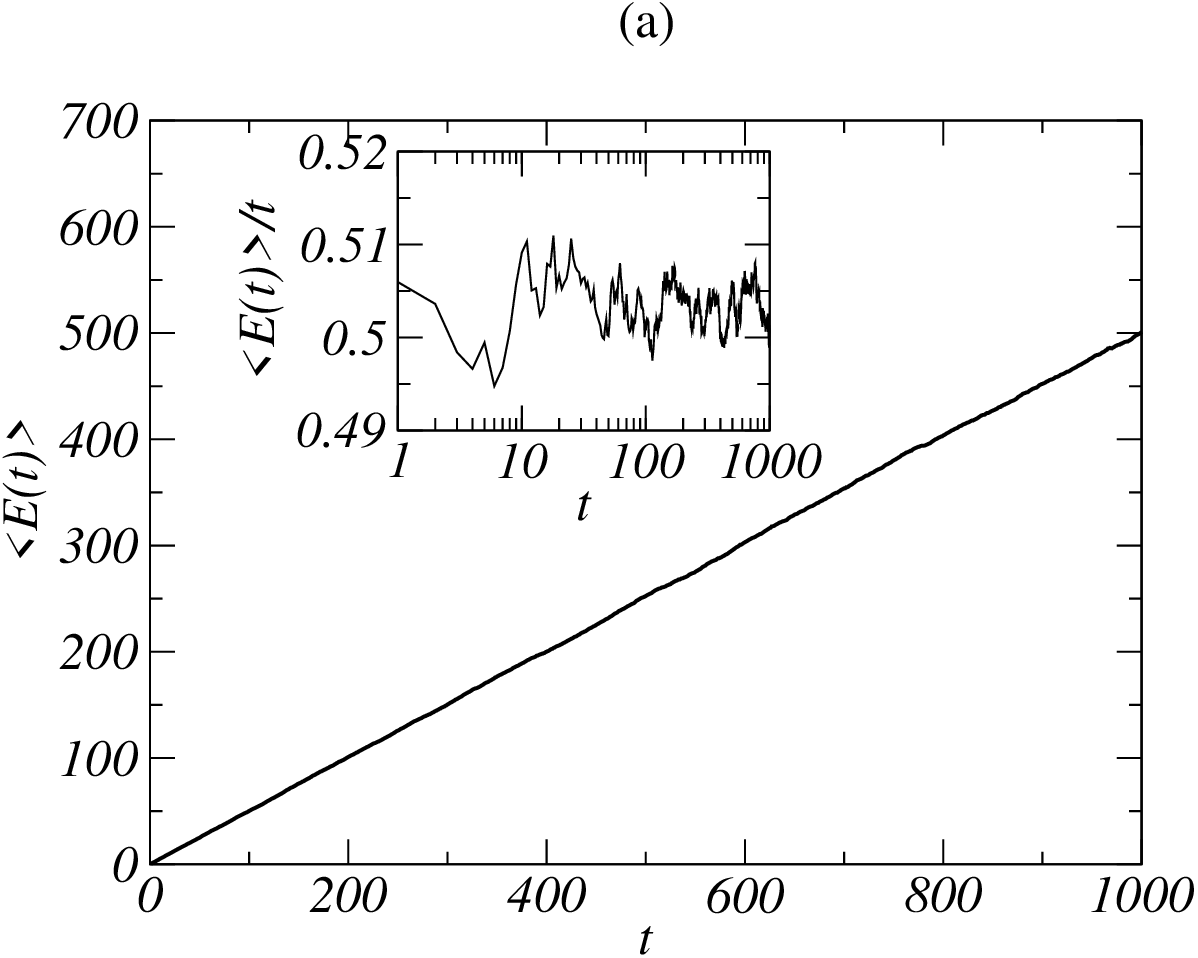}    
\hfill
\includegraphics*[width=0.4\textwidth]{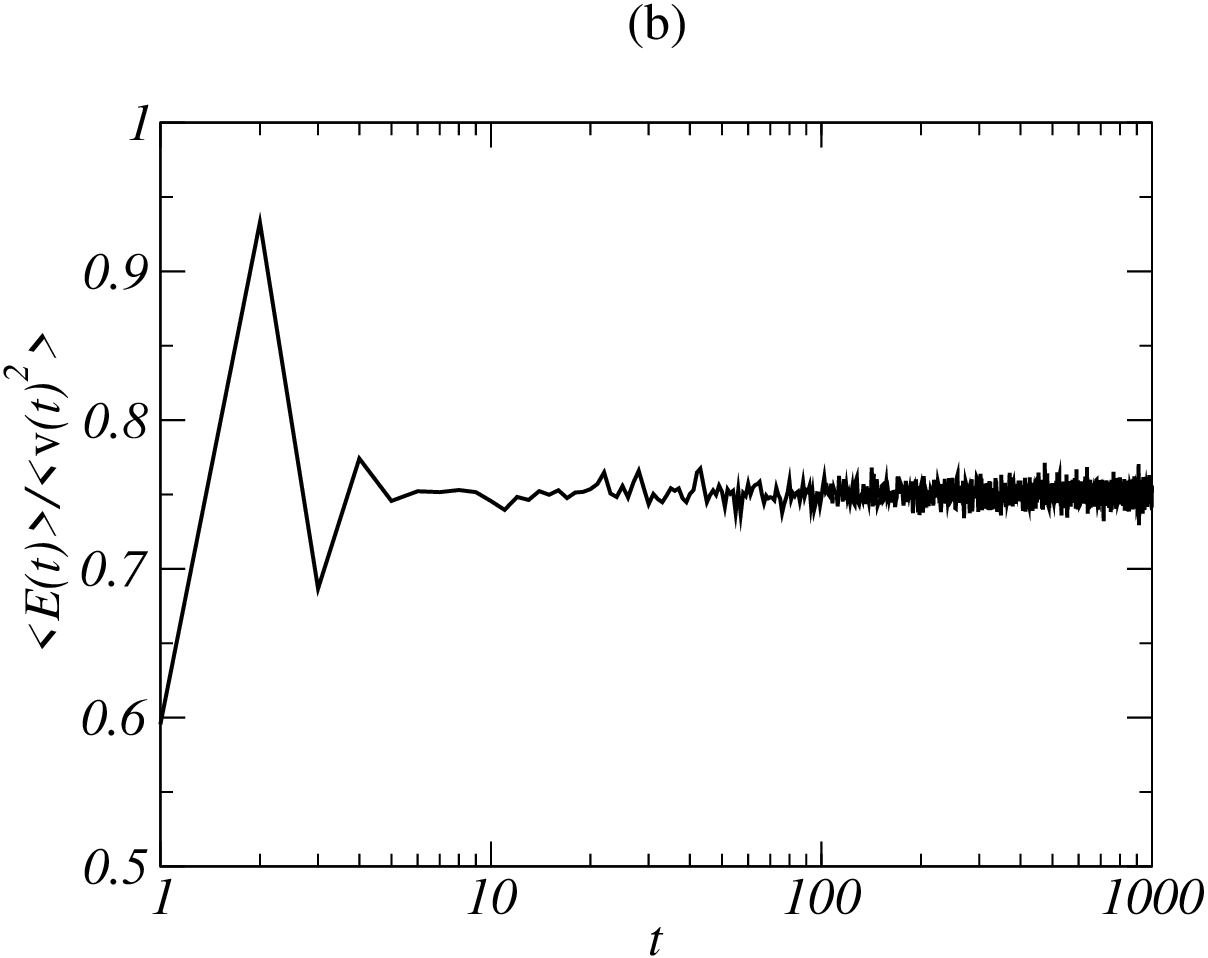}    
\hfill

    \caption{\label{fig:add:equi} Cubic oscillator with
additive noise.
Eq.~(\ref{addcubic}) is integrated numerically for ${\mathcal D} = 1$
with a timestep $\delta t = 5 \, 10^{-4}$. Ensemble averages are
computed over $10^4$ realizations.
Fig.~(a): the average energy $\langle E(t) \rangle$ grows 
linearly with time. The inset gives an estimate of the
diffusion constant when noise is additive:
$D_E \simeq 0.505(5)$. Fig.~(b): the time-asymptotic value
of the equipartition $\langle E(t) \rangle / \langle v(t)^2 \rangle$
is equal to $3/4$.
}
\end{figure}

When the   nonlinear  restoring  force  is cubic,  
  the linear term can be neglected in the long-time limit 
  and    the dynamics  is   given by
\begin{equation}
 \frac{\textrm{d}^2 }{\textrm{d} t^2}x(t) + x(t)^{3} =  \xi(t) \, ,
 \label{addcubic}
\end{equation}
where $\xi(t)$ is the Gaussian white noise defined in Eq.~(\ref{defgamma}).
This equation  can be analyzed  
 as in Sec.~\ref{sec:duff:deg}. After introducing 
 the  energy and angle variables, defined in Eqs.~(\ref{cubicE})
  and (\ref{cubicangle}), we obtain
\begin{eqnarray}
\dot E  &=& \dot x \, \xi(t) = (2E)^{\frac{1}{2}} \,
  \frac{ {\rm  cn}(\phi;{1}/{\sqrt{2}})   }
      { {\rm dn}^2(\phi;{1}/{\sqrt{2}})} \, \xi(t) ;
\label{evoladdE}     \\                              
 \dot\phi &=&  (4E)^{\frac{1}{4}} - \frac{ {\rm sd}(\phi;{1}/{\sqrt{2}})}
{ 2\sqrt{2}\, E^{1/2}} \, \xi(t)   .
   \label{evoladdphi}
\end{eqnarray}
 We  can  show that the  auxiliary variable $\Theta$  defined as
 \begin{equation}
  \Theta = E^{\frac{1}{2}}
\label{deftheta}    ,
\end{equation} 
 undergoes a normal diffusion process that can be precisely studied
   after averaging out 
  the fast variable $\phi$. The P.D.F.
  of the variable $\Theta$,  the scaling  laws
  for $E$, $\dot{x}$ and $x$,
 as well as  the generalized diffusion constants can be calculated
 exactly, as before. We  indicate here only  the asymptotic
  power-laws:
\begin{eqnarray}
          E      &\sim&  t ,  \nonumber   \\ 
          x      &\sim&  t^{\frac{1}{4} } , \nonumber    \\ 
         \dot x  &\sim&  t^{\frac{1}{2} }  .
\label{addscalings1}
\end{eqnarray}
 The  scaling exponents for additive noise are 
  different from those for 
  multiplicative noise. For exemple, we observe 
 that a particle submitted to an additive noise
 in a quartic potential is subdiffusive with an  anomalous
 exponent equal to $1/2$, whereas  in the presence of 
 multiplicative noise it behaves diffusively.

 In the general case, with a nonlinearity of the type $x^{2n -1}$,
 with $n \ge 1$,
 one  can prove  that 
 the   variable $\Theta$, defined in Eq.~(\ref{deftheta}),
 diffuses as   $t^{1/2}$ in the  long  time limit.
  Thus  the following  scalings 
   are satisfied  when $t \to \infty$:
\begin{eqnarray}
          E      &\sim&  t ,  \nonumber   \\ 
          x      &\sim&  t^{\frac{1}{2n}} , \nonumber    \\ 
         \dot x  &\sim&  t^{\frac{1}{2}  }  .
\label{addscalings2}
\end{eqnarray}
 All these exponents are smaller than the exponents found
 for a multiplicative white noise in Eq.~(\ref{scalings2}).

\hfill\break

 Finally,   when  both additive and
 multiplicative noises  are present, the  oscillator  is 
governed by the equation:
\begin{equation}
 \frac{\textrm{d}^2 }{\textrm{d} t^2}x(t) + x(t) \, \xi_{\textrm{mult}}(t)
+ x(t)^{2n-1} =  \xi_{\textrm{add}}(t) \, ,
 \label{addmult}
\end{equation}
 where  $\xi_{\textrm{mult}}$ and $\xi_{\textrm{add}}$ are independent 
white noises  of amplitude  ${\mathcal D}_{\textrm{mult}}$ and  
${\mathcal D}_{\textrm{add}}, $
 respectively. If we study  the energy variation 
 due to  noise,  $\dot E  \sim  x \dot x \xi_{\textrm{mult}} + 
\dot x \xi_{\textrm{add}}
  \simeq E^{\frac{n+1}{2n} } \xi_{\textrm{mult}} +  E^{1/2 } 
\xi_{\textrm{add}} $,
 we observe   that the  
  first term has a  dominant effect. 
 From this  simple   argument, we conclude that
  the multiplicative noise is expected to dominate over the additive
 noise and therefore, asymptotically,
 the scaling laws  will be  those derived  for  
 the multiplicative noise alone. However, 
 a  crossover between the two scalings
 (\ref{scalings2}) and (\ref{addscalings2}) should be observed 
 by choosing   ${\mathcal D}_{mult} \ll   {\mathcal D}_{add} $.
 Comparing Eqs.~(\ref{scalings2})  and  (\ref{addscalings2}), we find that
  the effect of the multiplicative noise  starts to dominate 
  after a crossover time of the order of
 $({\mathcal D}_{mult}\,t_c)^{\frac{1}{2n-2}} \sim
 ({\mathcal D}_{add}\,t_c)^{\frac{1}{2n}}   $ {\it i.e.}, 
 $ t_c \sim {\mathcal D}_{add}^{n-1}/{\mathcal D}_{mult}^n $.

  \section{Numerical Algorithm}
\label{sec:num}

The algorithm used to integrate numerically the stochastic 
ordinary differential equations studied in this article 
is the one-step collocation method advocated in \cite{mannella}.
In this Appendix, we recall the general principles
underlying this method, and give the algorithms we used
to integrate Eqs.~(\ref{nthorder})  and (\ref{colorn})-(\ref{OU}), for white
and correlated noise respectively. All stochastic equations
are understood according to Stratonovich rules.

Let $\{ x_i \}_{i=1,\ldots,N}$ be $N$ real variables of time $t$, 
and $\xi(t)$ a stochastic process  assumed to be Gaussian and white.
We wish to solve systems of $N$ coupled Langevin equations of the form
\begin{equation}
  \label{eq:lang:gen}
  \dot x_i = f_i(\{x_j(t)\}) + g_i(\{x_j(t)\}) \; \xi(t),
\end{equation}
where $f_i$ and $g_i$ are $N$ (smooth) functions of the $x_i$'s.
Let $\delta t$ be the integration timestep.
Upon integrating formally Eq.~(\ref{eq:lang:gen}) between $0$ and $\delta t$,
we obtain the following set of coupled equations implicit in $\{x_i(t)\}$
\begin{equation}
  \label{eq:lang:int}
  x_i(\delta t) - x_i(0) = \int_0^{\delta t} 
f_i(\{x_j(s)\}) \, \mathrm{d}s + \int_0^{\delta t}  
g_i(\{x_j(s)\}) \; \xi(s) \, \mathrm{d}s.
\end{equation}
For $\delta t$ small enough, the functions $f_i$ and $g_i$
may be Taylor-expanded in the vicinity of $t=0$, \emph{e.g.},
\begin{equation}
  \label{eq:Taylor}
  f_i(\{x_j(s)\}) = f_i(\{x_j(0)\}) + \partial_k f_i(\{x_j(0)\}) \,
\delta x_k(s) + \frac{1}{2} \, \left( \partial_k \partial_l f_i(\{x_j(0)\}) 
\right) \, \delta x_k(s) \, \delta x_l(s) + \ldots \, ,
\end{equation}
where $\delta x_k(s) = x_k(s) - x_k(0)$.
Replacing the functions $f_i$ and $g_i$ in Eq.~(\ref{eq:lang:int})
by the expansion (\ref{eq:Taylor}), we obtain a
new set of coupled integral equations implicit in $\{\delta x_i\}$.
Upon solving these equations up to an arbitrary
order in $\delta t$, we obtain $\delta x_k(\delta t)$.
In practice, we implemented an algorithm exact to $O(\delta t^2)$.

\subsection{White noise}
\label{sec:num:white}

We integrate the following set of first-order differential equations
\begin{eqnarray}
  \label{eq:white}
  \dot x &=& v,\\
  \dot v &=& - \omega^2 x - \lambda x^{2n-1} + x \, \xi,
\end{eqnarray}
where the Gaussian white noise $\xi(t)$ verifies Eq.~(\ref{defgamma}).
The exact evolution equations for $\delta x(\delta t)  = x(\delta t) - x(0)$
and $\delta v(\delta t)  = v(\delta t) - v(0)$ read
\begin{eqnarray}
  \label{eq:evol:white}
  \delta x(\delta t) &=& \int_0^{\delta t} \mathrm{d}s \,
\left( v(0) + \delta v(s) \right),\\
  \delta v(\delta t) &=& \int_0^{\delta t} \mathrm{d}s \,
\left\{ - \omega^2 \, \left( x(0) + \delta x(s) \right)
- \lambda \, \left( x(0) + \delta x(s) \right)^{2n-1}
+  \left( x(0) + \delta x(s) \right) \, \xi(s)
\right\}.
\end{eqnarray}
The auxiliary variables $Z_1$ and $Z_2$, defined by
\begin{eqnarray}
  Z_1(\delta t) &=& \int_0^{\delta t} \xi(s) \, \mathrm{d}s,\label{eq:defz1}\\
  Z_2(\delta t) &=& \int_0^{\delta t} Z_1(s) \, \mathrm{d}s,\label{eq:defz2}
\end{eqnarray}
are Gaussian random variables, with zero average, and
the following correlations:
$ \langle Z_1(\delta t)^2 \rangle = {\mathcal D} \, \delta t$,
$ \langle Z_2(\delta t)^2 \rangle = {\mathcal D} \, \delta t^3/3$,
$ \langle Z_1(\delta t) \, Z_2(\delta t) \rangle =
{\mathcal D} \, \delta t^2/2$.
Up to order $(\delta t)^2$, we find
\begin{eqnarray}
  \label{eq:algo:white}
  x(\delta t) &=& x(0) + v(0) \, \delta t
+ Z_2(\delta t) + \frac{1}{2} \, \delta t^2 \, 
\left( - \omega^2 \, x(0) - \lambda \, x(0)^{2n -1} \right),\\
  v(\delta t) &=& v(0) + Z_1(\delta t) +
 \delta t \, \left( - \omega^2 \, x(0) - \lambda \, x(0)^{2n -1} 
\right) + \frac{1}{2} \, \delta t^2 \, 
 v(0) \left( - \omega^2 \, x(0) -
(2n-1) \, \lambda \, x(0)^{2n-2} \right).
\end{eqnarray}
In practice, we use two {\em independent} Gaussian random noises 
$Z_1$ and $Y_1$ with zero mean and correlations 
$ \langle Z_1(\delta t)^2 \rangle = {\mathcal D} \, \delta t$,
$ \langle Y_1(\delta t)^2 \rangle = {\mathcal D} \, \delta t$.
As shown in Ref.~\cite{mannella}, the variable  $Z_2$ may be 
approximated by the expression:
\begin{equation}
  \label{eq:defz2y1}
  Z_2(\delta t) = \delta t \, \left( \frac{1}{2} Z_1(\delta t)
+ \frac{1}{2 \, \sqrt{3}} \, Y_1(\delta t) \right),
\end{equation}
when the algorithm is exact up to order $\delta t^2$.

\subsection{Colored noise}
\label{sec:num:colored}

When the noise $\eta(t)$ is correlated, we must  solve a set
of three coupled equations
\begin{eqnarray}
  \label{eq:color}
  \dot x &=& v,\\
  \dot v &=& - \omega^2 x - \lambda x^{2n-1} + x \, \eta,\\
  \dot \eta &=& - \frac{1}{\tau} \, \eta + \frac{1}{\tau} \, \xi.
\end{eqnarray}
The set of exact integral equations becomes
\begin{eqnarray}
  \label{eq:evol:color}
  \delta x(\delta t) &=& \int_0^{\delta t} \mathrm{d}s \,
\left( v(0) + \delta v(s) \right),\\
  \delta v(\delta t) &=& \int_0^{\delta t} \mathrm{d}s \,
\left\{ - \omega^2 \, \left( x(0) + \delta x(s) \right)
- \lambda \, \left( x(0) + \delta x(s) \right)^{2n-1}
+  \left( x(0) + \delta x(s) \right) \, 
 \left( \eta(0) + \delta \eta(s) \right) \right\},\\
  \delta \eta(\delta t) &=& \int_0^{\delta t} \mathrm{d}s \,
\left\{ - \frac{1}{\tau} \,  \left( \eta(0) + \delta \eta(s) \right)
+ \frac{1}{\tau} \, \xi(t) \right\}.
\end{eqnarray}
The algorithm to order $\delta t^2$ reads
\begin{eqnarray}
  \label{eq:algo:color}
  x(\delta t) &=& x(0) + v(0) \, \delta t
+ \frac{1}{2} \, v(0) \,\delta t^2 \, ,\\
  v(\delta t) &=& v(0) + 
 \delta t \, \left( - \omega^2 \, x(0) - \lambda \, x(0)^{2n -1}
 + x(0) \, \eta(0) \right) 
  + \frac{1}{\tau} \, x(0) \, Z_2(\delta t) \nonumber\\
   &&+ \frac{1}{2} \, \delta t^2 \, \left(
 v(0) \left( - \omega^2 \,  x(0) -
(2n-1) \, \lambda \, x(0)^{2n-2} + \eta(0) \right)
- \frac{1}{\tau} \, x(0) \, \eta(0)
\right),\\
  \eta(\delta t) &=& \eta(0) + \frac{1}{\tau} \, Z_1(\delta t)+
  - \frac{1}{\tau} \, \eta(0) \, \delta t
  - \frac{1}{\tau^2}  \, Z_2(\delta t)
+ \frac{1}{2} \, \delta t^2 \, \frac{\eta(0)}{\tau^2}.
\end{eqnarray}

   \section{Some useful Properties of Elliptic Functions}
\label{sec:elliptic}

 The Jacobian elliptic functions ${\rm sn}(u;k),\, {\rm cn}(u;k),\,$ and 
${\rm dn}(u;k)$  are meromorphic functions of the complex variable $u$. They
 depend on a parameter $k$ called the elliptic modulus \cite{abram,byrd}. These
 functions, like the trigonometric functions, have a real period
 and, like the hyperbolic functions,  have an imaginary period.
They are thus doubly periodic in the complex plane and their periods
 are $(4K, 2iK')$, $(4K, 2K+ 2iK')$ and $(2K, 4iK')$, respectively, where
\begin{equation}
    K(k) = \int_0^1 \frac{{\rm d}t}{\sqrt{( 1 -t^2)(1 - k^2t^2)}} \,\,\,
     = \int_0^{\frac{1}{k'}}  \frac{{\rm d}y}
  {\sqrt{( 1 -{k'}^2y^2)(1 +  k^2y^2)}} \,\,\, \hbox{ with } \,\,\,
 t^2 = \frac{y^2}{ 1 +k^2 y^2} \,\, \hbox{ and } \,\,\,
   {k'}^2 =  1 - k^2   \,\, .
 \end{equation}
  The imaginary period is given by
 $  K'(k) =  K(k') =  K\left(\sqrt{1 -  k^2}\right) \, .$
The Jacobian elliptic functions satisfy the fundamental relations:
\begin{eqnarray}
        {\rm sn}^2(\phi;k) + {\rm cn}^2(\phi;k)  &=&  1 \,,\nonumber \\
         k^2 {\rm sn}^2(\phi;k) + {\rm dn}^2(\phi;k)  &=&  1 \,,
       \label{carres} 
  \end{eqnarray}
  and their  derivatives are
 \begin{eqnarray}
 \frac{\textrm d}{{\textrm d}\phi}{\rm sn}(\phi;k)   &=& 
{\rm cn}(\phi;k)    \, {\rm dn} (\phi;k) \, , \nonumber \\
  \frac{\textrm d }{{\textrm d} \phi }{\rm cn}(\phi;k)     &=& 
-{\rm sn} (\phi;k)   \, {\rm dn}(\phi;k)   \,, \nonumber \\
    \frac{\textrm d }{{\textrm d}\phi}{\rm dn}(\phi;k)  &=& 
-k^2{\rm sn}(\phi;k) \, {\rm cn}(\phi;k) \,.
         \label{derivees}
\end{eqnarray}
 
  If we define the function  $ {\rm sd}(\phi;k) = 
{\rm sn} (\phi;k)  / {\rm dn} (\phi;k)\, , $  the relations
 (\ref{carres}) and (\ref{derivees}) lead to two new formulas
\begin{equation}
\frac{\textrm d}{{\textrm d}\phi}{\rm sd}(\phi;k) =
  \frac{ {\rm cn}(\phi;k)  }{  {\rm dn}^2(\phi;k) }
      \,\,\, \hbox{  and } \,\,\, 
  \frac{ {\rm cn}^2(\phi;k)  }{  {\rm dn}^4(\phi;k) }
 =  ( 1 - {k'}^2 {\rm sd}^2(\phi;k)) ( 1 + k^2 {\rm sd}^2(\phi;k)) \, .
 \label{reltsd}
\end{equation} 
Moreover, from  Eqs.~(\ref{derivees}) and (\ref{reltsd}), we obtain
\begin{eqnarray}
  {\rm sn}^{-1}(y;k) &=&
 \int_0^y \frac{{\textrm d}t}{\sqrt{( 1 -t^2)(1 - k^2t^2)}}  \, , 
\label{inversesn} \\
   {\rm sd}^{-1}(y;k) &=& \int_0^y \frac{{\textrm d}u}
  {\sqrt{( 1 -{k'}^2 u^2)(1 +  k^2 u^2)} }  \, .
   \label{inversesd}
\end{eqnarray}
 Similar relations can be derived  for ${\rm cn}^{-1}$ and ${\rm dn}^{-1}.$ 
 The expressions  of inverse elliptic functions in terms of
  quartic  integrals provide   a classic way of defining
  elliptic functions.  The change of variables
  $ t = u/\sqrt{1 +k^2 u^2}$ in Eq.~(\ref{inversesn})  leads to
  the identity 
\begin{equation} 
    {\rm sd}^{-1}(y;k) = {\rm sn}^{-1}
 \left( \frac{y}{ \sqrt{1 +k^2 y^2} }; k \right)
\label{liensnsd}
\end{equation} 
 We  deduce  from Eq.~(\ref{inversesn})
 \begin{equation}
   \frac{\textrm d}{{\textrm d} k}  {\rm sn}^{-1}(y;k) 
  =  k  \int_0^y  \frac{ t^2 {\textrm d}t}{\sqrt{( 1 -t^2)(1 - k^2t^2)}}
  = k  \int_0^{y/\sqrt{1 - k^2 y^2}}  \frac{  u^2{\textrm d}u}
  {\sqrt{( 1 -{k'}^2 u^2)(1 +  k^2 u^2)} }
  = k \int_0^{\phi_1} {\rm sd}^2(\phi;k) {\textrm d}\phi  \, ,
\label{derivparam}
\end{equation} 
 where the  last equality was obtained by setting $ u = {\rm sd}(\phi;k) $
 with $ \phi_1 = {\rm sn}^{-1}(y;k) \, .$ 

\hfill\break

 We  verify from Eqs.~(\ref{carres}) and (\ref{derivees}) 
 that $x$ and $\dot{x}$ defined in Eqs.~(\ref{solx}) and (\ref{solv})
 are solutions of the cubic oscillator $\ddot{x} + x^3 =0.$
 Similarly, we  verify  
  that $x$ and $\dot{x}$ defined in Eqs.~(\ref{dufx}) and 
(\ref{dufv})  satisfy  the differential equation for the Duffing oscillator,
 $ \ddot{x} + \omega^2 x +  x^3 =0.$  We now 
  prove the identities  used to derive  
  the equipartition relations and the diffusion  constants
 in section \ref{sec:duff:deg}. In the case of a  purely
 cubic oscillator,  $ k = k' = 1/{\sqrt 2} \, ,$
 and   we obtain from  Eq.~(\ref{inversesd}) 
\begin{equation}
  \frac{\textrm d}{{\textrm d} y} {\rm sd}^{-1}(y;\frac{1}{\sqrt{2}}) 
 = \frac{1}{\sqrt{ 1 - \frac{y^2}{4}}} \,.
\label{invsdrac2}
\end{equation}
 From Eq.~(\ref{reltsd}) we deduce 
 \begin{equation}
\overline {\frac{ {\rm cn^2}(\phi)} { {\rm dn}^4(\phi) }     }
  = \frac{1}{K} \int_0^K {\rm d}\phi \,
  \frac{ {\rm cn^2}(\phi)} { {\rm dn}^4(\phi) } = \frac
{  \int_0^1  \sqrt{ 1 - u^4}  {\textrm d}u   }
 {  \int_0^1 \frac {{\textrm d}u}{  \sqrt{ 1 - u^4}} } = \frac{2}{3}  \, ,
 \label{moycn2}
\end{equation}
 where we have set $ u = {\rm sd}(\phi)/\sqrt{2}$ to obtain the second
 equality;  the final equality results from Eq.~(\ref{ipp}).
Similarly, we obtain:
\begin{equation}
\overline {\frac{ {\rm cn^4}(\phi)} { {\rm dn}^8(\phi) }     } =
  \frac{1}{K} \int_0^K {\rm d}\phi \,
  \frac{ {\rm cn^4}(\phi)} { {\rm dn}^8(\phi) } 
   = \frac {  \int_0^1  ( 1 - u^4)^{\frac{3}{2}}  {\textrm d}u  }
 {  \int_0^1 \frac{ {\textrm d}u}{  \sqrt{ 1 - u^4}} }
 = \frac{4}{7}  \,  .
\label{moycn4dn8}
\end{equation}
 The last  equality is obtained  by the following integration  by parts:
 $$ \int_0^1  1. ( 1 - u^4)^{\frac{3}{2}} \, {\textrm d}u 
 = \int_0^1 6 u^4 ( 1 - u^4)^{\frac{1}{2}} \, {\textrm d}u 
 = \int_0^1 6 ( 1 - u^4)^{\frac{1}{2}} \, {\textrm d}u 
- \int_0^1 6 ( 1 - u^4)^{\frac{3}{2}} \, {\textrm d}u  \, ;$$ 
 thus
$$ \int_0^1   ( 1 - u^4)^{\frac{3}{2}}  {\textrm d}u 
  = \frac{6}{7} \int_0^1  ( 1 - u^4)^{\frac{1}{2}} {\textrm d}u \, .$$
 Now,  using Eq.~(\ref{moycn2}),  we obtain the
 numerical value 4/7 in Eq.~(\ref{moycn4dn8}).
 We also   have:
\begin{equation}
 \overline{{\rm sd^2}(\phi)} = 
\frac{1}{K} \int_0^K {\rm d}\phi \, {\rm sd^2}(\phi)  = 2 
    \frac {  \int_0^1 \frac{ u^2 {\textrm d}u}{  \sqrt{ 1 - u^4}} }
    {  \int_0^1 \frac{ {\textrm d}u}{  \sqrt{ 1 - u^4}} } =
\frac{5}{2} {\mathcal M},
\end{equation}
 where ${\mathcal M}$ was defined in Eq. (\ref{defM}). The last
 equality is obtained by using the following identities
 $$ \int_0^1 \frac{  u^6 {\textrm d}u}{  \sqrt{ 1 - u^4} }
  = \int_0^1 \frac{u^3}{2} \frac{ 2 u^3 {\textrm d}u}{  \sqrt{ 1 - u^4} }
   = \frac{3}{2}  \int_0^1 u^2 \sqrt{ 1 - u^4} {\textrm d}u  \, , $$
 $$  \int_0^1 \frac{ u^2 {\textrm d}u}{  \sqrt{ 1 - u^4}}{\textrm d}u
 = \int_0^1 \left(  u^2 \sqrt{ 1 - u^4}   + 
   \frac{u^6} {  \sqrt{ 1 - u^4} }    \right){\textrm d}u 
  = \frac{5}{2} \int_0^1   u^2 \sqrt{ 1 - u^4}   {\textrm d}u     \, . $$

 We end this  Appendix by explaining how  the equations
  for the phase  dynamics are obtained.  Eq.~(\ref{evolangle})
 is derived from  Eqs.~(\ref{invsdrac2}) and (\ref{evolE}):
 \begin{equation}
   \frac{ {\textrm d}}{{\textrm d} t} \phi
 =  \frac{ {\textrm d}}{{\textrm d} t} 
{\rm sd}^{-1}\left(\frac{x}{E^{1/4}}, \frac{1}{\sqrt{2}}\right)
 = 
 \frac{  \frac{ \dot x}{E^{1/4}} -  \frac{  x \dot E}{4 E^{5/4}}}
 {\sqrt{ 1 - \frac{x^4}{4E}}} = 
 \left( \frac{ \dot x}{E^{1/4}} -  \frac{  x^2  \dot x \xi}{4 E^{5/4}} \right)
 \frac{ \sqrt{2 E}}{ \dot x} =(4E)^{\frac{1}{4}} -
  \frac{ {\rm sd}^2(\phi;{1}/{\sqrt{2}})}
{ 2\sqrt{2} E^{1/4}} \xi(t) \,   .  
\end{equation}
 In order to derive the dynamics of the phase
  of a Duffing oscillator  (\ref{duffangle}),
  we must differentiate  Eq.~(\ref{cubicangle2})
 with respect to time.  We must bear in mind  that the elliptic
 modulus $k$  defined in Eq.~(\ref{modulus}) is now
 a function of time. Therefore, when  differentiating  the
 function ${\rm sd}^{-1}$,  we obtain two contributions:
  $(i)$  the derivative  with respect to the  argument produces a result
 similar to the one for the cubic case,
 $(ii)$  the derivative with respect to the elliptic modulus, calculated
 using  Eqs.~(\ref{liensnsd}) and (\ref{derivparam}),
  generates the  two last terms  in Eq.~(\ref{duffangle}).

\end{document}